\documentclass[aps,prb,12pt,onecolumn,citeautoscript]{revtex4-1} 
\usepackage{amsmath,amssymb,mathrsfs,bm,feynmf,setspace,soul}
\usepackage{graphicx}
\usepackage[tight]{subfigure}  
\usepackage{color}   
\usepackage[papersize={8.5in,11in}]{geometry}
\usepackage[colorlinks=true]{hyperref} 
\hypersetup{
    bookmarks=true,         
    unicode=false,          
    pdftoolbar=true,        
    pdfmenubar=true,        
    pdffitwindow=false,     
    pdfstartview={FitH},    
    pdftitle={My title},    
    pdfauthor={Author},     
    pdfsubject={Subject},   
    pdfcreator={Creator},   
    pdfproducer={Producer}, 
    pdfkeywords={keyword1} {key2} {key3}, 
    pdfnewwindow=true,      
    colorlinks=true,       
    linkcolor=red,          
    citecolor=blue,        
    filecolor=magenta,      
    urlcolor=cyan           
} 
\newcommand{\de}{\delta}

\newcommand{\la}{\lambda}

\newcommand{\s}{\sigma}

\newcommand{\D}{\Delta}

\newcommand{\ket}[1]{\left| #1\right\rangle}

\newcommand{\beq}{\begin{equation}}
\newcommand{\eeq}{\end{equation}}
\newcommand{\Beq}{\begin{eqnarray}}
\newcommand{\Eeq}{\end{eqnarray}}
\newcommand{\bml}{\begin{multline}}

\newcommand{\eeqm}{\end{multline}}

\newcommand{\bsp}{\begin{split}}
\newcommand{\esp}{\end{split}}
\newcommand{\down}{\downarrow}
\newcommand{\up}{\uparrow}

\renewcommand{\b}[1]{{\bm #1}}

\newcommand{\mc}{\mathcal}

\newcommand{\req}[1]{Eq.~(\ref{eq:#1})}

\newcommand{\rfig}[1]{Figure~\ref{fig:#1}}
\newcommand{\nn}{\nonumber}

\begin{document}

\title{Correlated quantum phenomena in the strong spin-orbit regime} 
 \author{William Witczak-Krempa}
 \affiliation{Perimeter Institute for Theoretical Physics, Waterloo, Ontario N2L 2Y5, Canada}
 \author{Gang Chen}
 \affiliation{Department of Physics, University of Colorado, Boulder, Colorado, US}
 \author{Yong Baek Kim}
 \affiliation{Department of Physics, University of Toronto, Toronto, Ontario M5S 1A7, Canada}
 \affiliation{School of Physics, Korea Institute for Advanced Study, Seoul 130-722, Korea} 
 \author{Leon Balents}
 \affiliation{Kavli Institute of Theoretical Physics, University of California, Santa Barbara, Santa Barbara, CA, 93106}
\date{\today}
\begin{abstract}
  We discuss phenomena arising from the combined influence of
    electron correlation and spin-orbit coupling, with an emphasis on
    emergent quantum phases and transitions in heavy transition metal
    compounds with 4$d$ and 5$d$ elements.  A common theme is the
    influence of spin-orbital entanglement produced by spin-orbit
    coupling, which influences the electronic and magnetic structure.  In
    the weak-to-intermediate correlation regime, we show how non-trivial
    band-like topology leads to a plethora of phases related to
    topological insulators.  We expound these ideas using the example
    of pyrochlore iridates, showing how many novel phases such as the
    Weyl semi-metal, axion insulator, topological Mott insulator, and
    topological insulators may arise in this context.  In the strong
    correlation regime, we argue that spin-orbital entanglement fully or
    partially removes orbital degeneracy, reducing or avoiding the
    normally ubiquitous Jahn-Teller effect. As we illustrate for the
    honeycomb lattice iridates and double perovskites, this leads to
    enhanced quantum fluctuations of the spin-orbital entangled states
    and the chance to promote exotic quantum spin liquid and multi-polar
    ordered ground states.  Connections to experiments,
    materials, and future directions are discussed.

\vspace{1cm}

\end{abstract} 
\maketitle
\tableofcontents

\section{Introduction}

The subject of this review is the combination of two central threads
of quantum materials research.  The first, correlated electron
physics, is a venerable but still vibrant subject, born from
observations of Mott, Hubbard, Anderson, and others on the properties
of $3d$ transition metal oxides.  It is largely concerned with the
diverse properties of electronic materials which are insulating, or in
the process of becoming so, as a result of electron-electron
interactions \cite{mott,RevModPhys.70.1039}, most importantly the strong
local Hubbard repulsion $U$ between electrons occupying the same
orbital.  A plethora of phenomena arises from correlated electron physics, including 
local moment formation and magnetism, correlated metallic states,
quantum criticality, and unconventional
superconductivity\cite{RevModPhys.70.1039}.  The second thread,
non-trivial physics from strong spin-orbit coupling (SOC), includes a
body of work on $f$-electron materials\cite{RevModPhys.81.807} and the much more recent activity
begun with the theoretical proposal of topological insulators (TIs) in
2005\cite{RMP_TI,RMP_TI2,moore-hasan}.  SOC is a relativistic effect,
which provides an interaction between the orbital angular momentum and
electron spin in atoms, and is usually considered a small perturbation
in the discussion of electrons in solid.  However, in heavy elements
it need not be weak -- it effectively increases proportionally to
$Z^4$, where $Z$ is the atomic number -- and indeed has striking
qualitative effects.  Since 2005, the investigation of topological
aspects of electron bands has exploded, both theoretically and
experimentally\cite{RMP_TI,RMP_TI2,moore-hasan}.  From the materials
perspective, the domain of the TI field has mostly been the class of solids
with heavy $s$- and $p$-electron elements such as Bi, Pb, Sb, Hg, and Te.
In these materials, topologically protected Dirac-like surface states
have been predicted and observed, and a host of further phenomena are
currently under intense investigation.

The two research strands come together in the heavy transition metal compounds
drawn especially from the $5d$ series, and in some cases the $4d$'s as
well.  Upon descending the periodic table from the $3d$ to $4d$ to the $5d$
series, there are several competing trends.  First, the $d$ orbitals
become more extended, tending to reduce the electronic repulsion $U$ and thereby
diminish correlation effects.  However, simultaneously, the SOC
increases dramatically, leading to enhanced splittings between otherwise
degenerate or nearly degenerate orbitals and bands, reducing in many
cases the kinetic energy.  The latter effect can offset the reduction in
$U$, allowing correlation physics to come into play.  

\begin{figure}[b]
\includegraphics[width=12cm]{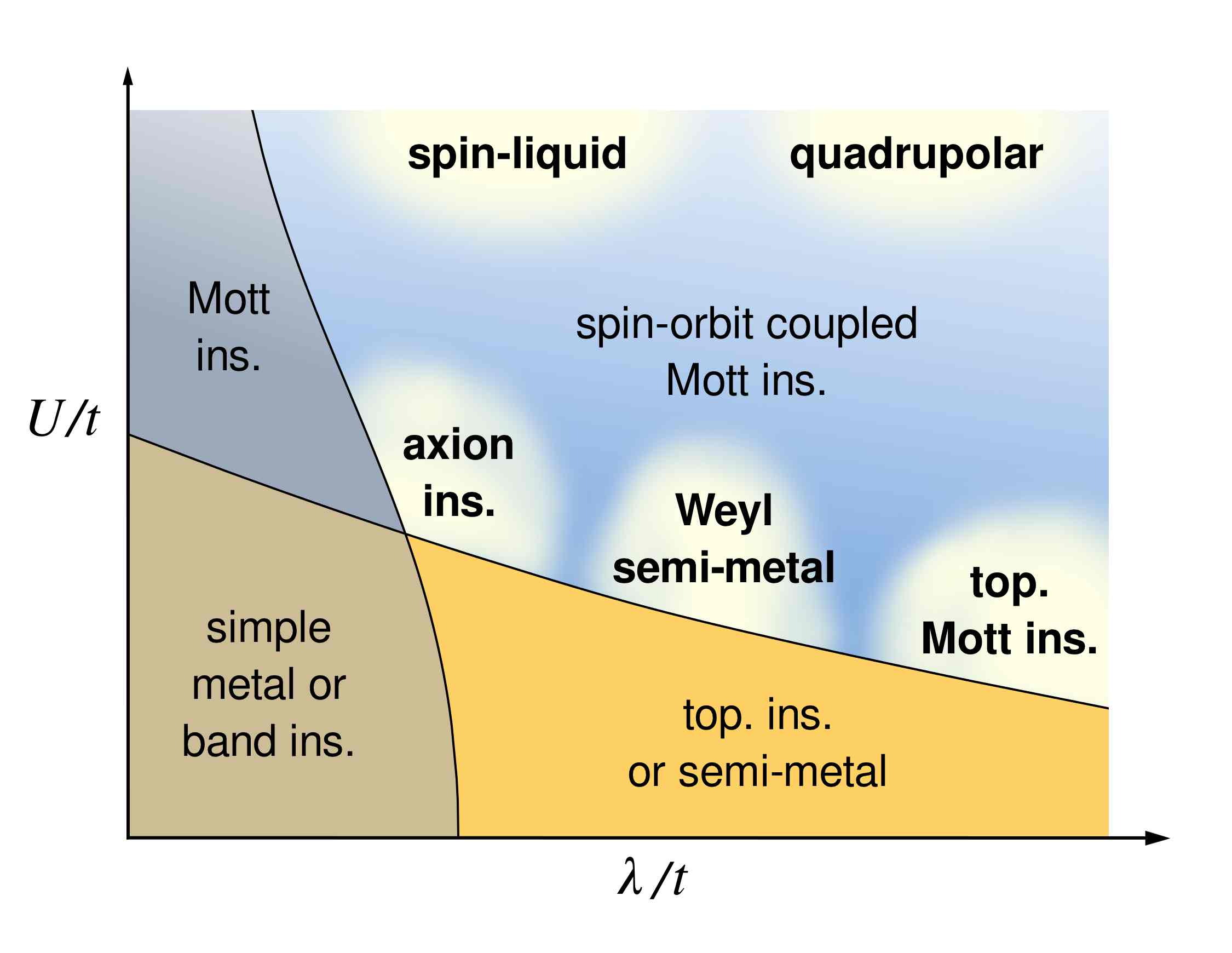}
\caption{
Sketch of a generic phase diagram for electronic materials, in terms of the
interaction strength $U/t$ and SOC $\lambda/t$. The materials in this
review reside on the right half of the figure.
}
\label{fig:gen-pd}
\end{figure}

It is instructive to consider a generic model Hamiltonian describing the above discussion:
\begin{equation}
H = \sum_{i,j; \alpha \beta} t_{ij,\alpha\beta} \ c^{\dagger}_{i \alpha} c^{\vphantom\dagger}_{j \beta} + {\rm h.c.}
+ \lambda \sum_i  {\b L}_i \cdot {\b S}_i + U \sum_{i,\alpha}
n_{i\alpha} (n_{i\alpha} -1) \ ,
\label{eq:5}
\end{equation}
where $c_{i \alpha}$ in the annihilation operator for an electron in orbital $\alpha$ at site
$i$ and $n_{i\alpha}=c^\dagger_{i\alpha}c^{\vphantom\dagger}_{i\alpha}$, the corresponding
occupation number, $t$ the hopping amplitude, $\lambda$ the atomic SOC
entangling spin ($\b S_i$) and angular momentum ($\b L_i$),
and  $U$ the Hubbard repulsion.  
An explicit example of the spin-orbital entanglement due to $\la$ is given later in \req{J_eff}.  
We have for simplicity omitted the Hund’s interaction between spins in different orbitals on the same site, 
which is much smaller than $U$ but can sometimes have significant effects\cite{rev-georges} -– it is, however, unimportant in the specific examples discussed in detail in this review.
A schematic ``phase diagram'' can be drawn as in
\rfig{gen-pd} in terms of the relative strength of the interaction $U/t$ and the
SOC $\lambda/t$\cite{pesin}.  We emphasize this is schematic, in part 
because the problem is unsolved, and in part because Eq.~\eqref{eq:5}
can represent many different physical situations by the choice of
orbitals and lattice/band structure encoded in $t_{ij,\alpha\beta}$, and
the ground states which occur certainly depend upon these choices.  In
this diagram, two lines (which are not meant necessarily as sharp
boundaries but rather as demarcating different limits) divide the weak
and strong correlation regions, and the weak and strong SOC regimes,
thereby generating four quadrants.  Conventional transition metal materials
reside on the left hand side of the diagram, where SOC is weak $\lambda/
t \ll 1$, and a conventional metal-insulator transition (MIT) may occur when
$U$ is comparable to the bandwidth, or a few times $t$. Upon increasing
SOC, when $U/t\ll 1$, a metallic or semi-conducting state at small $U$
may be converted to a semi-metal or to a TI.  What happens when both SOC
and correlations are present?  Several arguments suggest that $\lambda$
and $U$ tend to cooperate rather than compete, in generating insulating
states.  Including SOC first, we have already remarked upon the
splitting of degeneracies and the consequent generation of multiple
narrow bands from relatively mixed ones.  The narrow bands generated by
SOC are more susceptible to Mott localization by $U$, which implies that
the horizontal boundary in \rfig{gen-pd} shifts
downward with increasing $\lambda$.  If we include
correlations first, the $U$ tends to localize electrons, diminishing
their kinetic energy.  Consequently the on-site SOC $\lambda$, which is
insensitive to or even reduced by \emph{delocalization}, is relatively
enhanced.  Indeed, in the strong  Mott regime $U/t \gg 1$,  one should compare $\lambda$
with the spin exchange coupling $J \propto t^2/U$, rather than $t$.   As a
result, the vertical boundary shifts to the left for large $U/t$.   We
see that there is an intermediate regime in which insulating states are
obtained only from the combined influence of SOC and correlations -- 
these may be considered {\em spin-orbit assisted Mott insulators}.  Here
we are using the term ``Mott insulator'' to denote any state which is
insulating by virtue of electron-electron interactions.  In Sec.~\ref{sec:concl}, we
will remark briefly on a somewhat philosophical debate as to what should
``properly'' be called a Mott insulator.   

Terminology aside, an increasing number of experimental systems have
appeared in recent years in this interesting correlated SOC regime.
Most prolific are a collection of {\em iridates}, weakly conducting or
insulating oxides containing iridium, primarily in the Ir$^{4+}$
oxidation state.  This includes a Ruddlesdon-Popper sequence of
pseudo-cubic and planar perovskites
Sr$_{n+1}$Ir$_{n}$O$_{3n+1}$ ($n=1,2,\infty$)\cite{SrIrO-prl,SrIrO-science,takagi214_2012,yjkim214_2012,bjkim214_2012,bjkim327_2012,takagi327_2012,takagi327_2013}, hexagonal insulators
(Na/Li)$_2$IrO$_3$\cite{PhysRevB.82.064412,
gegenwart213_2012,yjkim213_2013,yjkim213_2011,PhysRevB.86.195131,PhysRevLett.108.127204},
a large family of pyrochlores,
$R_2$Ir$_2$O$_7$\cite{maeno,takagi-mit,Bi_Cao12}, and some spinel-related
structures\cite{takagi_hyperkagome,Ir2O4}.  Close to
iridates in the periodic table are several osmium oxides such as
NaOsO$_3$\cite{PhysRevB.80.161104} and
Cd$_2$Os$_2$O$_7$\cite{PhysRevLett.108.247205}, which experimentally display MITs.  Apart from these
materials where the microscopic SOC is truly rather strong ($\lambda
\approx$ 0.4eV), there are also examples which arise in materials
further ``up'' in Figure~\ref{fig:gen-pd}, in which electrons 
are strongly localized, the competing exchange scale is anomalously
weak, making even smaller SOC ``strong''.  This includes the
spin-orbital liquid material FeSc$_2$S$_4$\cite{loidl_2005,loidl_2006,chen2008sos}, possibly the single layer vanadate Sr$_2$VO$_4$\cite{jackeli09},
and a host of double perovskites, $A_2BB'$O$_6$\cite{Cussen06,Vries10,Aharen10,Carlo11,Qu13,Vries13,Aharen102,Wiebe03,Wiebe02,yamamura06,Stitzer02,Erickson07,Steele11,Aharen09}, where the exchange scale is small because
the electronically active $B'$ ions are well separated spatially.  For
space reasons, we cannot discuss all of these materials, but will take
examples from this list as illustrations.   

This review is organized based on a consideration of two limits within
this domain: the weak-to-intermediate correlation regime, $U/W
\lesssim 1$, where $W$ is the bandwidth which is typically a few times $t$,
and the strong Mott limit, $U/W \gg 1$, both in the
presence of strong SOC.  Conceptually, in the former, the electrons
remain delocalized enough that band or band-like topology continues to
play an important role, as in the case of
TIs\cite{RMP_TI,RMP_TI2,moore-hasan}.  Notably, interactions open
up possibilities for new types of topological phases.  This includes
Weyl semi-metals (WSM) with Dirac-like bulk quasiparticles and surface
``Fermi-arc" states\cite{volovik-book,wan,will-pyro1,burkov}, or axion insulators
with unusual electromagnetic response\cite{wan12,ara-pyro1}, which can
arise in the presence of spontaneous magnetic order.  Phase
transitions between these states\cite{pesin,wan,will-pyro1,burkov,wan12,ara-pyro1,bj-trig}, and amongst metals and insulators,
also occur in this regime, and yet more exotic phases have also been
envisioned\cite{pesin,will-tmi}.  We discuss this weak to intermediate correlation regime
in Sec.~\ref{sec:weak-interm-corr}, taking the pyrochlore iridates,
$R_2$Ir$_2$O$_7$, as primary experimental examples. 

In the strong Mott regime, discussed in Sec.~\ref{strongU}, electron band
topology no longer plays a role, since electronic states are not extended.
However, SOC still offers new physics by fully or partially lifting 
orbital degeneracy of partially filled $d$ shells, not by
ordering orbitals, but by {\em entangling} the orbital and spin
degrees of freedom\cite{chen2008sos,chen10,chen11}.  This provides a distinct mechanism to avoid the
Jahn-Teller effect and classical orbital ordering.  The orbital
degeneracy may be fully lifted by SOC, which occurs in iridates in
the strong Mott+SOC limit, or partially lifted, which is the case
for many $d^1$ or $d^2$ ions.  In this review we illustrate both
possibilities, taking the honeycomb
iridates\cite{PhysRevB.82.064412,
gegenwart213_2012,yjkim213_2013,yjkim213_2011,PhysRevB.86.195131,PhysRevLett.108.127204}, Na$_2$IrO$_3$ and Li$_2$IrO$_3$, as
examples of the former, and double perovskites\cite{Cussen06,Vries10,Aharen10,Carlo11,Qu13,Vries13,Aharen102,Wiebe03,Wiebe02,yamamura06,Stitzer02,Erickson07,Steele11,Aharen09}, as examples of the latter.  In either case, the strong SOC results in strongly
anisotropic exchange interactions, and for the case of partial
degeneracy lifting, these have a highly non-trivial {\em multi-polar} nature\cite{chen10,chen11}.
We describe how these unusual interactions can promote large quantum fluctuations
and lead to novel quantum ground states that are not possible without
strong SOC\cite{chen10,chen11,Dodds11}. In particular, quantum spin liquid and quadrupolar
(spin nematic) ordered phases have been suggested to occur in these
systems.  

\begin{table}
  \caption{Emergent quantum phases in correlated spin-orbit coupled
    materials.  All phases have U(1) particle-conservation symmetry --
    \emph{i.e.}\ superconductivity is not included. Abbreviations are as
    follows: TME = topological
    magnetoelectric effect, TRS = time reversal symmetry, P =
    inversion (parity), (F)QHE = (fractional) quantum Hall effect,
    LAB = Luttinger-Abrikosov-Beneslavskii\cite{moon2012non}.
    Correlations are W-I = weak-intermediate,
    I = intermediate (requiring magnetic order, say, but mean field-like), and S = strong.
    $[A/B]$ in a material's designation signifies a heterostructure with alternating $A$ and $B$ elements.
  }\label{tbl:phases1}
\begin{ruledtabular}
\begin{tabular}{ p{4.0cm} | p{1.7cm} c p{5.8cm} | p{2.0cm}}
	\bf{Phase} & \bf{Symm.} & \bf{Correlation} & \bf{Properties} & \bf{Proposed materials} \\ \hline
\hline\hline
Topological Insulator & TRS & W-I & Bulk gap, TME, protected $\qquad$ surface
states & many \vspace{1 mm} \\ \hline
Axion Insulator & P & I & Magnetic insulator, TME, $\qquad$ no protected
surface states & $R_2$Ir$_2$O$_7$, $A_2$Os$_2$O$_7$\vspace{1 mm} \\ \hline
WSM & Not both TRS \& P & W-I & Dirac-like bulk states, 
surface Fermi arcs, anomalous Hall  & $R_2$Ir$_2$O$_7$,
HgCr$_2$Se$_4$, \ldots\vspace{1 mm} \\ \hline
LAB Semi-metal & cubic + TRS & W-I & Non-Fermi liquid &
$R_2$Ir$_2$O$_7$ \vspace{1mm} \\ \hline
Chern Insulator  & broken TRS & I & Bulk gap, QHE 
& Sr[Ir/Ti]O$_3$, $R_2[B/B']_2$O$_7$ \vspace{0.2mm} \\ \hline\hline
Fractional Chern $\quad$ Insulator  & broken TRS & I-S & Bulk gap, FQHE & Sr[Ir/Ti]O$_3$
\vspace{1mm} \\ \hline
Fractional Topological Insulator, Topological Mott Insulator & TRS & I-S &
Several possible phases.  Charge gap, fractional excitations &
$R_2$Ir$_2$O$_7$ \vspace{1mm} \\ \hline
Quantum spin liquid & any & S & Several possible phases.  Charge gap,
fractional excitations & (Na,Li)$_2$IrO$_3$, Ba$_2$YMoO$_6$ \vspace{1mm}
\\ \hline
Multipolar order & various & S & Suppressed or zero magnetic moments.
Exotic order parameters. & $A_2BB'$O$_6$\vspace{1mm}
\end{tabular}
\end{ruledtabular}
\end{table} 

Table~\ref{tbl:phases1} summarizes the emergent phases encountered in
this review.   The theoretical and experimental status of the
weak-intermediate and strong correlation  regimes are presented in
Sections~\ref{sec:weak-interm-corr} and \ref{strongU}, respectively.  Following
this, we conclude the review with a discussion of open issues and
other materials in Sec.~\ref{sec:concl}.

\section{Weak to intermediate correlations}
\label{sec:weak-interm-corr}

Following the discovery of topological insulators\cite{RMP_TI,RMP_TI2} (TIs), it is now
recognized that SOC is an essential ingredient 
in forming certain topological phases.  The TI is characterized by a
$\mathbb Z_2$ topological invariant, which may be obtained from the
band structure, in the presence of time-reversal symmetry (TRS).  This phase
has a bulk excitation gap, but is distinguished from trivial band
insulators by the presence of protected metallic surface states. The
non-trivial bands in the TI must ``unwind'' upon crossing the
interface with a time-reversal preserving vacuum or a trivial band
insulator, leading to the closing of the excitation gap and a
conducting state at the interface.  Stemming as it does from band
structure considerations, the TI relies upon some degree of
itineracy.  Other TI-related phases with gapless nodes -- point Fermi
surfaces -- in the bulk have also been explored theoretically, and
obviously also require delocalized electronic states.  

In this section, we discuss phenomena involving band-like topology in
the presence of interactions.  Non-trivial band topology can probably only arise
when correlations are not so strong as to localize electrons fully to
single atoms.  Consequently, we focus on this regime of weak to
intermediate correlations.  From the theoretical point of view, this
means that a Hubbard model, rather than one of localized spins, is
likely a better starting point.   

What new phenomena can be expected, relative to the uncorrelated $s$- and
$p$-electron materials which are the mainstay of TI experiments?  TIs
obviously are stable to interactions, and quantitatively, correlations
may even increase the gap in some cases (With correlations,
in contrast to the free case, the surface states have the potential of spontaneously
breaking TRS or even exhibiting exotic fractionalization\cite{senthil-class,*metlitski,*fidkowski,*bonderson}). 
A more qualitatively novel prospect is to probe topological phases with spontaneous time-reversal 
breaking, since magnetism is common in correlated materials.  In
general, the $\mathbb Z_2$ classification fails for time-reversal
broken systems, and instead Chern insulators, \emph{i.e.}\ materials with
quantized Hall effects, may occur for instance.  In the presence of crystalline
symmetries, notably inversion, a $\mathbb Z_2$ index may 
reappear\cite{mong10,wan,turner-inv,hugues-inv,Fang13,kargarian13}.  
This is the case in the ``axion insulator''\cite{wan,wan,turner-inv,hugues-inv}, which may be
characterized by a quantized magnetoelectric effect, {\it i.e.}\ an
electric polarization $\b P$ can be generated by applying a magnetic
field $\b B$: $\b P = \frac{\theta}{(2\pi)^2}\b B$ with $\theta = \pi$
such that the ratio $P/B$ is universal and quantized.  In fact the
same is true for three-dimensional TIs, and the quantized
magnetoelectric effect can be used to define TIs\cite{Wang-prl10}
and axion insulators\cite{Wang2012,wang_simplified,ara-pyro1} in the presence of interactions. 
 
\begin{figure}
\centering
\subfigure[]{\label{fig:weyl}\includegraphics[scale=.33]{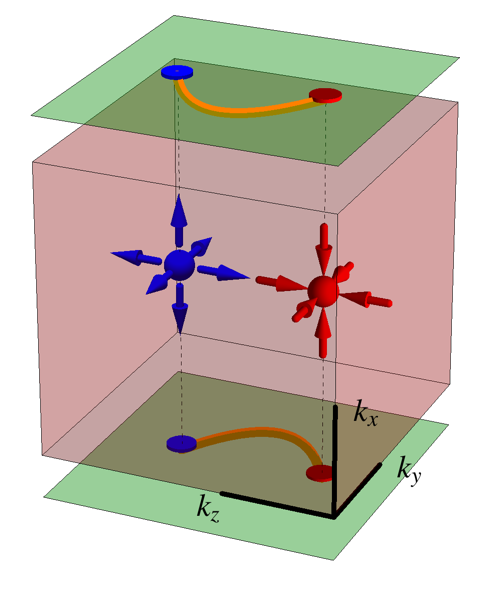}}  
\caption{\label{fig:nnn-bs} 
Two opposite-chirality Weyl points in the three dimensional  
Brillouin zone (BZ), and the associated Fermi arc
surface states. The green sheets correspond to the top and bottom surface-BZs.
This minimal WSM must break TRS, and will have a corresponding anomalous
quantum Hall conductivity, $\s_{xy}=(e^2/2\pi h)\D k$, where $\D k$ is the momentum-space
separation between the two Weyl points. 
}
\end{figure}

Non-trivial topology can also appear in gapless phases, such as in the
WSM\cite{volovik-book,murukami,wan}.  At the level of band theory, the WSM has a Fermi
surface consisting of points, where only two bands meet linearly, see \rfig{nnn-bs}. This
implies that either TRS or inversion is broken, for otherwise all bands
would be two-fold degenerate.   In this review, we will encounter examples
arising from a spontaneous breaking of TRS at sufficiently large $U$.
The band touchings of the WSM are three-dimensional analogs of the
massless Dirac fermions in clean graphene.  However, contrary to the
latter, they should be regarded as topological objects -- monopoles or 
hedgehogs in momentum space -- because they act as sources of the
momentum space Berry flux\cite{volovik-book,ahe_haldane,murukami,wan}. Consequently, they always come in pairs with opposite 
chiralities, corresponding to a positive or negative monopole charge, as illustrated in
\rfig{nnn-bs}. An
isotropic example of a Weyl fermion is given by the following Bloch
Hamiltonian (we take the touching to be at $\b k_W$)
\begin{align} \label{eq:weyl}
  \mc H(\b k) = \pm v (\de k_x\s_x+\de k_y\s_y+\de k_z\s_z)\,, \quad \de\b k=\b k-\b k_W
\end{align}
where $\pm$ corresponds to positive/negative chirality and $\s_i$ are
the Pauli matrices acting on the space of the two non-degenerate bands
touching at the Weyl point.  Because any local perturbation is a
combination of Pauli matrices, it can only move the touching in the Brillouin zone (BZ)
and not lift it, since all $\s_i$ already appear in $\mc H$.  As a
consequence of the non-trivial bulk topology, the WSM hosts
non-trivial surface states on certain boundaries, which take the form
of open Fermi arcs\cite{wan}, as shown in \rfig{weyl}. This cannot
occur in a purely 2D system, just like the existence of an odd number of
Dirac fermions at the surface of a three-dimensional $\mathbb Z_2$ TI,
which apparently violates the ``fermion doubling'' theorem\cite{doubling}. In both
cases, however, the ``partner" electronic states exist on the opposite
surface, which elegantly resolves the apparent paradox.  For some
further discussion of WSMs, including other realizations,\cite{burkov,dai_fang11,halasz} and of
generalizations of these ideas to superconductors,\cite{meng13} we
refer readers to recent reviews.\cite{turner2013beyond,vafek-rev}

The Chern insulator, axion insulator, and WSM require interactions to
produce magnetic order, but once produced, a mean field description in
terms of a static exchange field suffices, and the electronic
quasiparticles are consequently weakly correlated.  Other phases arising
in interacting systems with strong SOC may be more intrinsically
correlated.  Several interacting analogs of TIs have been envisioned,
which are not adiabatically connected to the ground states of {\em any}
non-interacting electron Hamiltonian, with or without broken symmetries.
These include the topological Mott insulator of Ref.~\onlinecite{pesin}, which
exhibits {\em spin-charge separation} and TI-like surface states
composed of neutral fermions, and {\em fractional} Chern insulators\cite{sheng11,neupert11,tang} 
which display a fractional quantum Hall effect without an external magnetic field.  Presently, while we
believe the {\em ingredients} for these sorts of phases are present in
the class of materials discussed in this review, a direct
connection of specific compounds to specific states of this type is an
open theoretical challenge.  

A final category of phenomena to be mentioned in the intermediate
correlation regime are thermal and quantum {\em phase transitions}.
The elephant in the room is the MIT itself,
whose character in the strong SOC limit has been very little
investigated, but may be very different from what has been observed in
traditional $3d$ compounds.  For example, theory has established that
the MIT may occur through an intermediate WSM
state, and transitions to/from WSM states may be studied.  Other types
of correlation-driven MITs have been suggested.
Note that at $T>0$, the difference between a metal and an insulator is
quantitative not qualitative (since $\sigma \neq 0$ always), so that
the evolution from a metal to an insulator at $T>0$ needs not coincide with
any phase transition. The onset of magnetic order in correlated metals
and insulators constitutes another type of criticality.  Such
transitions may again be affected by strong SOC.

\subsection{Pyrochlore iridates}

In this section we expand on the theory and experimental
status of new phases and transitions in the context of the
pyrochlore iridates, $R_2$Ir$_2$O$_7$ ($R$-227) where $R$ is a rare earth
element.  We choose these compounds because they comprise a large
family for which experiments have revealed thermal phase transitions
and an evolution from metallic to insulating states, and largely
systematic variation of properties with $R$.  Many key theoretical
ideas in the field have also been introduced in this context,
including topological Mott insulators\cite{pesin,will-tmi,kargarian11}, chiral
spin liquids\cite{nakatsuji-pr}, WSMs\cite{wan,will-pyro1}, and axion
insulators\cite{ara-pyro1,wan}.  This has contributed to a substantial
experimental
effort\cite{maeno,taira,maeno-Y-doped,takagi-mit,takagi-mit-full,raman,
  nakatsuji2011musr,nakatsujiEu,takagi-nd,julian,Nd-pressure,
  fisher,disseler-magOrder-nd,disseler-magOrder-y-yb,nakatsuji-RXD}, 
in particular to determine the nature of the elusive magnetic
ordering.  We shall first review the experimental knowledge of the
pyrochlore iridates, and will then describe the theoretical work they
have motivated.

\begin{figure}
\centering 
\subfigure[]{\label{fig:latt} \includegraphics[scale=0.49]{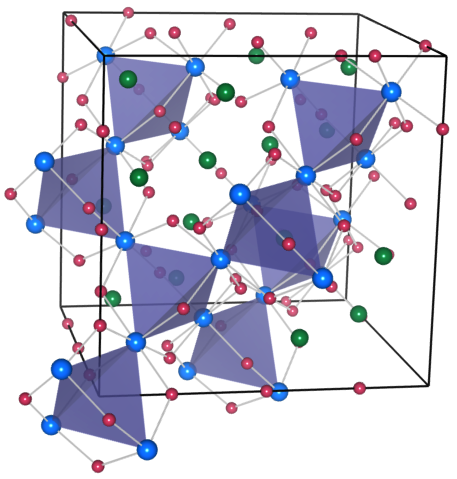}}
\subfigure[]{\label{fig:trig} \includegraphics[scale=0.38]{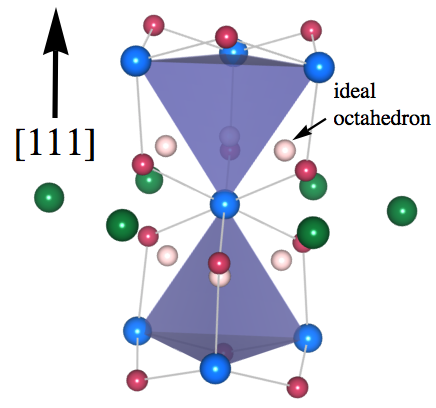}}
\caption{\label{fig:pyro-xtal} Pyrochlore lattice structure showing the Ir (blue) tetrahedra. Trigonally
compressed cages of oxgyens (small \& red) surround each Ir. The rare earth $R$ ions are in green.
The box is aligned along the [100], [010] and [001] directions. In b), the light spheres would
correspond to the position of oxygen ions in an ideal octahedron,
expected for a larger $R^{3+}$ ion. All pyrochlore iridates show trigonal compression of the
oxygen cages along the [111] directions. }
\end{figure}

\subsubsection{Experimental resume}
\label{sec:pyro-exp}
The structure of these materials is well established, consisting of
interpenetrating pyrochlore nets for both the rare earth and iridium ions\cite{subramanian,maeno}.
Each such lattice can be described as a face centered cubic (fcc) Bravais lattice with
four sites per unit cell forming a tetrahedron. The crystal structure is illustrated in \rfig{pyro-xtal}.
The fcc lattice constant was found to range
from $a=1.01$ to $1.04$~nm\cite{maeno,raman,fisher}, depending on the $R$ element. Another important structural parameter
is the oxygen $x$-parameter,
which determines the symmetry of the oxygen environment around
each Ir cation: when $x=x_{\rm ideal}=5/16=0.3125$ the oxygens form a perfect octahedron around 
each Ir.
Otherwise, the oxygen octahedron becomes elongated ($x<x_{\rm ideal}$) or compressed ($x>x_{\rm ideal}$) along
the $[111]$ directions.
These distortions are called trigonal because they preserve three-fold rotational symmetries.
Experiments show that the pyrochlore iridates all have compressed octahedra\cite{taira,fisher}. 

\begin{figure} 
\centering%
 \subfigure[]{\label{fig:eu-data} \includegraphics[scale=.37]{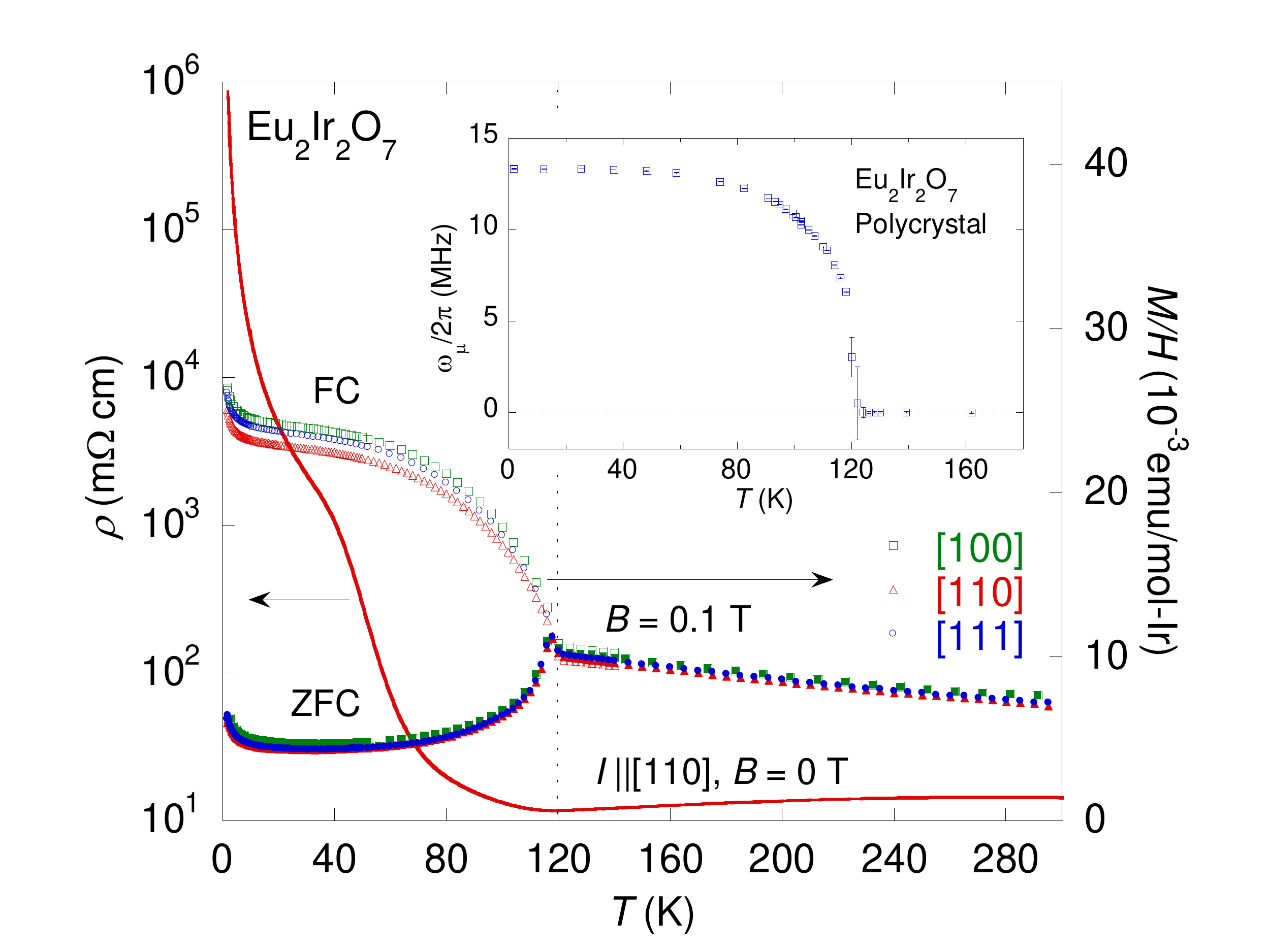}}
 \subfigure[]{\label{fig:pyro-pd} \includegraphics[scale=0.26]{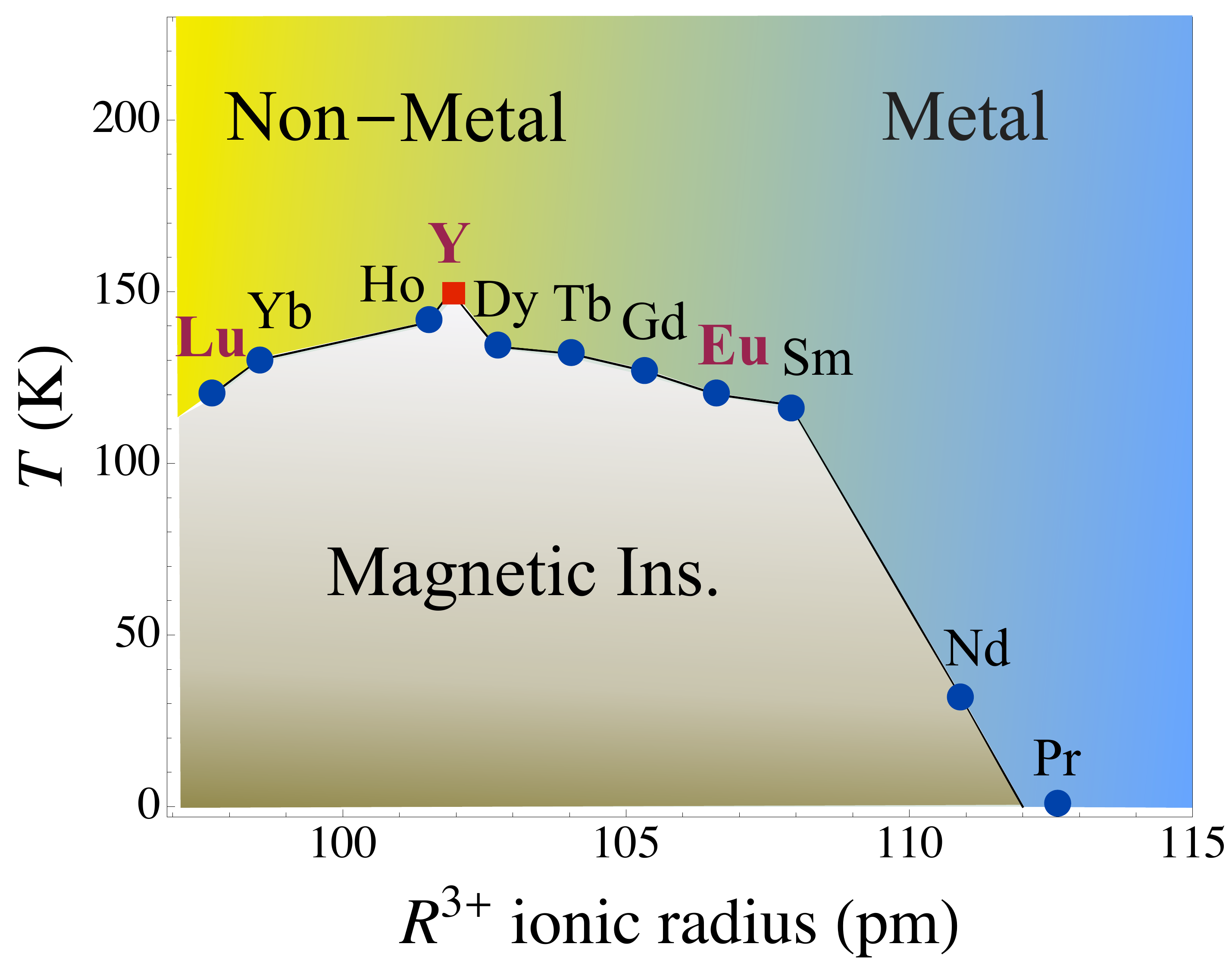}}
\caption{\label{fig:takagi} {a) The main panel shows the resistivity, and
field cooled (FC) and zero field cooled (ZFC) susceptibilities for Eu-227, while the insert
the spontaneous muon oscillation frequency. Data adapted
from Refs.~\onlinecite{nakatsujiEu,nakatsuji2011musr}.
b) Phase diagram for the pyrochlore iridates $R$-227 based on transport and magnetism
measurements. (This is a supplemented and modified version of the diagram found in Ref.~\onlinecite{takagi-mit-full}.) 
The $R$-elements that do not have a local magnetic moment are emphasized in bold magenta.
The only non-lanthanide, $R=$~Y, is denoted by a square.}  
}  
\label{iridate_data}
\end{figure}

Let us first review the temperature
dependence of the resistivity, $\rho$. All members of the family show a diverging $\rho$ at
sufficiently low temperature, except Pr-227\cite{maeno,takagi-mit,takagi-mit-full}. 
For the compounds with a larger $R^{3+}$ ion such
as Eu-, Sm- and Nd-227 the resistivity goes from being ``metallic'' ($d\rho/dT>0$)
at large temperatures $T>T_c$ to ``non-metallic'' ($d\rho/dT<0$) at smaller temperatures $T<T_c$. 
This is illustrated for Eu-227 in \rfig{eu-data}.
Pr-227 lies at the end of the spectrum in that it has the
largest ionic radius, and does not show a major resistivity upturn down to the lowest 
temperatures, in agreement with the above trend.
For compounds with smaller $R^{3+}$ ions such as for $R=$ Lu, Yb, Ho, Y, etc the slope does not change sign and the resistivity is
``non-metallic'' throughout. In that case, the upturn at $T_c$ is smoother. One can tentatively explain this trend according to 
which a larger R$^{3+}$ cation
leads to a more metallic $\rho$ as follows: Larger $R^{3+}$ ions lead
to a decreased trigonal compression of the octahedra (\rfig{trig}), which increases the Ir-O orbital-overlap and thus facilitates
the hopping of the Ir electrons\cite{koo}.

A variety of sharp features in the magnetic properties occur at the same
temperature as the resistivity upturn, $T_c$, which support its
interpretation as a true phase transition.  Most notably, the field
cooled (FC) and zero-field cooled (ZFC) magnetic susceptibilities branch
away from each other\cite{taira,takagi-mit}, as shown in \rfig{eu-data} for Eu-227.
In addition, $\mu$SR experiments, which have been performed on Eu-\cite{nakatsuji2011musr}, Nd-\cite{disseler-magOrder-nd,guo_Nd}  
Yb-\cite{disseler-magOrder-y-yb}, and Y-227\cite{disseler-magOrder-y-yb}, 
show the continuous rise of 
a well-defined muon-precession frequency directly below $T_c$ (illustrated in \rfig{eu-data} for Eu-227).  
This is indicative of long-range magnetic ordering into a commensurate 
structure. However, due to difficulties in performing neutron scattering experiments on 
Ir-based compounds, the precise nature of the ordering remains unknown.
Nevertheless, it has been inferred that a {\em second-order} phase transition to an AF insulator occurs at $T_c$.
Ferromagnetism can be largely ruled out by the absence of magnetic hysteresis\cite{maeno,taira}.
The second order nature of the transition is further supported by the presence of peaks in the  
heat capacity at $T_c$\cite{takagi-mit,takagi-mit-full}. The essence of the experimental
observations on the pyrochlore iridates is summarized in the phase diagram \rfig{pyro-pd}.

\subsubsection{Electronic structure}
\label{pyro}

The basic element of the description of these materials is the iridium
electronic structure.  We will neglect for the most part the rare
earth magnetism, which plays a role in some (but not all) of the
compounds at very low temperature.  Following the analysis of the
Ir-electron physics, we will briefly examine their interplay with
local moments in Section~\ref{sec:fd-exchange}.

We begin by examining the local atomic physics associated with the Ir
cations. The outer-shell electrons of Ir$^{4+}$ are in a 5$d^5$
configuration, half-filling the ten $d$-levels. The dominant crystal field
splitting comes from the oxygen octahedra surrounding each Ir cation,
which splits the levels into a higher energy $e_g$ orbital doublet and
a lower $t_{2g}$ orbital triplet, spanned by orbitals with $xy, yz$ and $zx$ symmetry. 
These are separated by a $\sim
2$ eV gap and as such we can neglect the higher energy $e_g$ levels.
We then must take into account both the SOC and trigonal distortions.
Let us begin with atomic SOC in Ir$^{4+}$ ions.  The full angular
momentum operator ${\b L}$ projected to the $t_{2g}$ manifold acts as
an effective angular momentum one, ${\b L}_{\rm eff}$, up to a minus
sign, {\it i.e.}\ ${\cal P}_{t2g} {\b L}{\cal P}_{t2g} = - {\b L}_{\rm eff}$, where ${\cal P}_{t2g}$ represents the projection on
the $t_{2g}$ manifold.  Therefore, the SOC $\lambda$ splits the
$t_{2g}$ spinful manifold into a higher energy $J_{\rm eff}=1/2$ doublet and a
lower $J_{\rm eff}=3/2$ quadruplet. The states in these multiplets exhibit spin-orbital entanglement. 
For example, the Jeff = 1/2 states are
\begin{align}\label{eq:J_eff}
  \ket{J_{\rm eff}^z=+1/2} &= \frac{1}{\sqrt{3}}(\ket{xy,\up}+\ket{yz,\down}+i\ket{zx,\down})\,, \nn\\
  \ket{J_{\rm eff}^z=-1/2} &= \frac{1}{\sqrt{3}}(-\ket{xy,\down}+\ket{yz,\up}-i\ket{zx,\up})\,. 
\end{align}
In an ionic picture, since
Ir$^{4+}$ has 5 $d$-electrons, the $J_{\rm eff}=1/2$ doublet is
half-filled, and only this orbital is involved in the low energy
electronic structure.  More generally, if trigonal splitting is
included, the $J_{\rm eff}=3/2$ levels are split and mixed with the
$J_{\rm eff}=1/2$ ones.  In the general case, there is a highest Kramers
doublet, whose character varies with the ratio of SOC to trigonal
splitting, between a $J_{\rm eff}=1/2$ doublet and a $S=1/2$ one.  

A band structure view is complementary to the ionic picture as we now discuss.  
If only the highest doublet is
involved, we expect 4 two-fold degenerate bands near the Fermi energy,
as there are 4 Ir per unit cell.  As discussed by Wan {\em et al.}\cite{wan}
and Yang {\em et al.}\cite{bj-trig}, it
is instructive to consider their structure at the $\Gamma$ point.  Due
to cubic symmetry, the 8 Bloch states at this point decompose into 2
two-dimensional irreducible representations (irreps) and 1
four-dimensional irrep.  By electron counting, these bands should be
half-filled, so that if the order of these irreps, in terms of
degeneracies, is 2-2-4 or 4-2-2, a band insulating state may occur,
while if the order is 2-4-2, the 4-dimensional irrep must be
half-filled and hence the system cannot be gapped at the band
structure level (see the lowest panel of \rfig{nnn-bs-tsig-08}). The former situation was obtained by
Ref.~\onlinecite{pesin} based on a phenomenological but ad-hoc Hubbard
model for small $U$.  They found a transition from a
semi-metallic ground state to a TI one with increasing the ratio of SOC to
hopping. 
Subsequently, by {\em ab initio} methods, Wan {\em et al.}\
found\cite{wan} the latter, 2-4-2, ordering of irreps in Y-227.  In this case, a
TI is impossible, but other topological phases can occur with
increasing correlations.  For those iridates with metallic
paramagnetic states, the 2-4-2 scenario may be more plausible. 

A convenient tight-binding model which covers both limits was
introduced in Ref.~\onlinecite{will-pyro1,krempa_go13}.  It considers the minimal number of degrees of
freedom, {\it i.e.}\ a single Kramers doublet per site, which leads to a
total of 8 bands.  One can show that the most general nearest-neighbor
(NN) TRS
Hamiltonian has the form\cite{imada}
\begin{equation}\label{eq:Hgen}
H_{0}= \sum_{\substack{
    \langle i, j\rangle }} c_{i}^\dag(t_1+i t_2\b d_{ij}\cdot\b \s) c_{j} \,,
\end{equation} 
where the hopping parameters $t_1$ and $t_2$ are real, and $\b\s$ is a
vector of Pauli matrices acting on the pseudospin degree of freedom.
The $t_2$ term generates non-trivial Berry phases for the
hopping electrons, and will play a key role in realizing topological
phases.  The real vector $\b d_{ij}$ is aligned along the opposite
bond of the tetrahedron containing $i,j$.  
Diagonalization of $H_0$ reveals a semi-metallic state with the 2-4-2
ordering when $-2\leq t_2/t_1 \leq 0$, and a TI otherwise\cite{imada,krempa_go13}.  Notably,
the semi-metallic state so obtained is not a compensated system with
electron-hole pockets but a {\em zero gap semiconductor}.  More
recently it was argued that such a state forms a stable {\em non-Fermi
  liquid} phase with a quadratic band touching at the $\Gamma$ point,
christened a ``Luttinger-Abrikosov-Beneslavskii'' (LAB) phase for
historical reasons.\cite{moon2012non}

\subsubsection{Magnetism and Weyl fermions}

\begin{figure}
\centering%
\subfigure[]{\includegraphics[scale=.42]{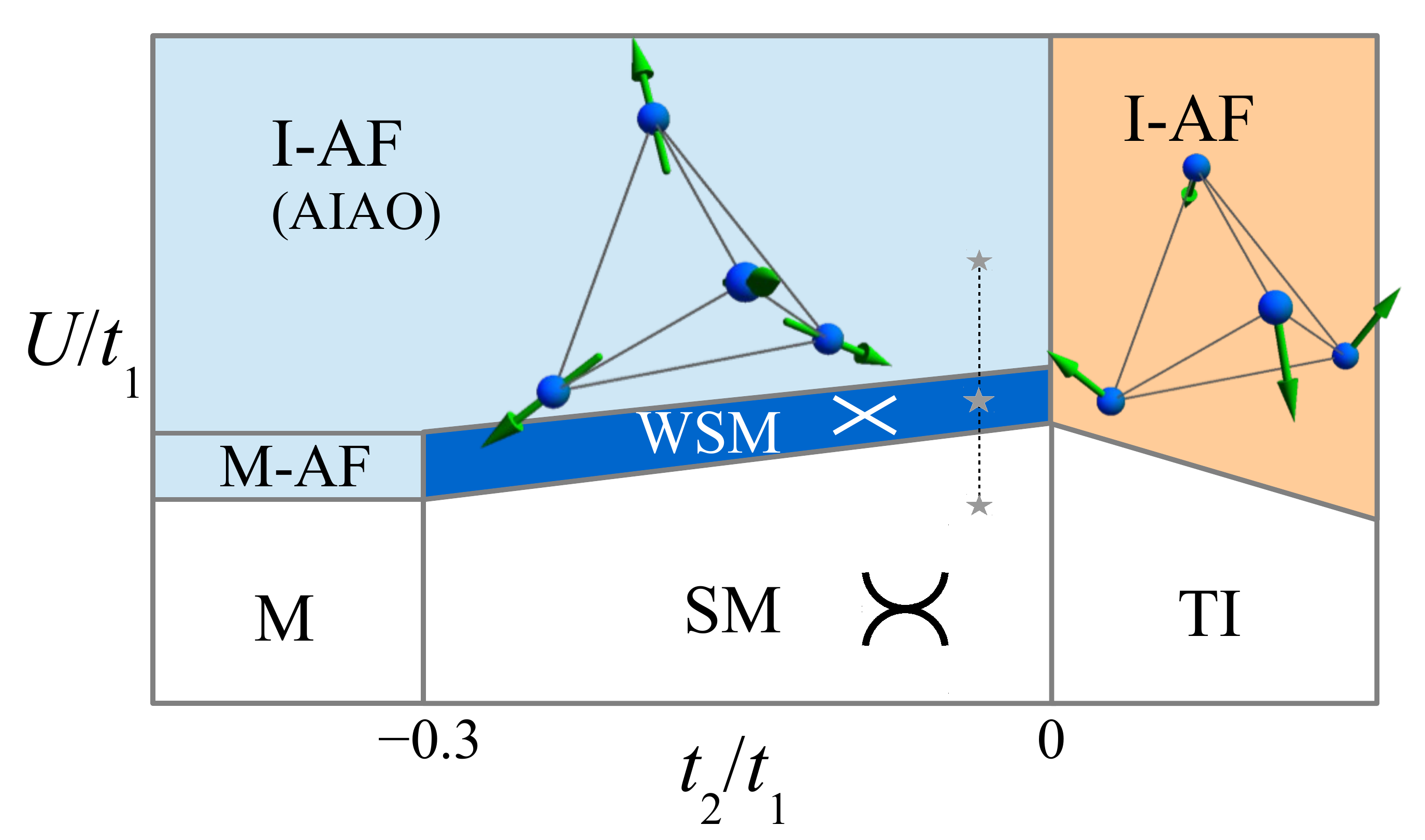}}
\subfigure[]{\label{fig:nnn-bs-tsig-08}\includegraphics[scale=.48]{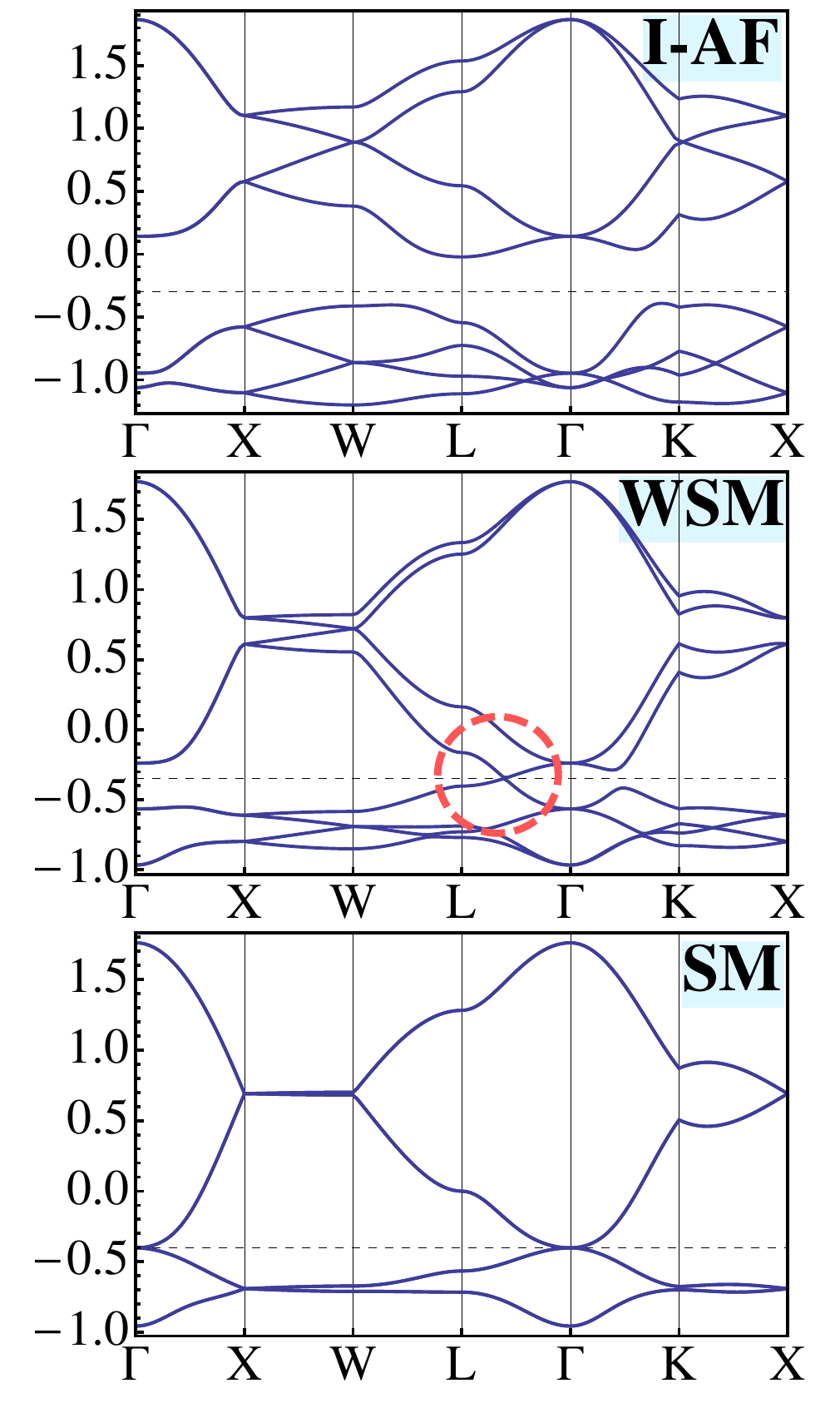}}  
\caption{\label{fig:hf} a) Schematic mean-field phase diagram of the Hubbard model on the
pyrochlore lattice\cite{will-pyro1,krempa_go13} (the kinetic Hamiltonian, \req{Hgen}, has been supplemented by small next nearest neighbor
  hopping). The shaded regions harbor antiferromagnetic (AF) order, which
can be of the AIAO type (blue) or a related type (orange). M = Metal, I/M-AF= Insulating/Metallic AF.
b) The evolution of the electronic spectrum along the vertical dashed line in a). 
The red circle shows one Weyl point at the Fermi level (horizontal dashed line).   
}
\end{figure}

As discussed in Sec.~\ref{sec:pyro-exp}, the experimental results strongly suggest antiferromagnetic order in
the low-temperature insulating state of the iridates.  This is also indicated from
theory.  LDA+U calculations by Wan {\em et al.}\ found an antiferromagnetic (AF) ground state
with each spin oriented along its local trigonal axis and all spins on
a given tetrahedron pointing in or out of that tetrahedron\cite{wan}.  Such an
``all-in/all-out'' (AIAO) state, illustrated in \rfig{hf}, maintains the unit cell and cubic symmetry of
the lattice, and is consistent with the lack of any indication of
structural transitions in experiment.  The same AIAO state
was obtained in a phenomenological Hubbard model, by theoretical
methods of varying sophistication (Refs.~\onlinecite{will-pyro1,ara-pyro1,krempa_go13,ara-pyro2}).   

In the magnetically ordered state, the broken TRS but preserved
inversion symmetry admits the possibility of both WSM and axion
insulator phases. Wan {\em et al.}\ indeed found a WSM phase for a
range of $U$ in their LDA+U calculations, and suggested but did not find
an axion insulator state\cite{wan}.  In fact, introduction of
arbitrarily weak AIAO magnetic order converts the quadratic
band touching LAB phase to a Weyl
semi-metal\cite{will-pyro1,krempa_go13}, which is illustrated in the middle panel of \rfig{nnn-bs-tsig-08}.  
As remarked earlier, this
phase is stable and indeed the Weyl points must migrate from the  
$\Gamma$ point toward the zone boundary to pairwise annihilate before
a true insulator is found at larger $U$.  Such behavior was found by
adding a Hubbard repulsion $U$ to the tight-binding model of
Eq.~\eqref{eq:Hgen}, within a Hartree-Fock
approximation\cite{will-pyro1,krempa_go13} (small second neighbor
hopping is needed to obtain the generic linear dispersion relation in
the WSM phase with AIAO order).  This is illustrated in Figure~\ref{fig:hf}.  The
quantitative width of the WSM in phase space varies between different
calculations, and can be very narrow, owing to rather flat bands and
the relatively fast growth of the local moment as the AIAO phase is
entered.  Other types of magnetic order, \emph{e.g.} ferromagnetism, can also
induce Weyl points, or relatives of them.\cite{will-unpub,moon2012non} 

Many signatures of the WSM have been suggested.  We have already
mentioned the surface Fermi arcs.  Without disorder, the low frequency
optical conductivity scales as $\sigma(\omega) \sim \omega$.  Though a pair
of opposite Weyl points mediates an intrinsic Hall conductivity\cite{ahe_haldane,ran}, this 
sums to zero given the cubic symmetry of the AIAO state,
which dictates at least 8 nodes.  However, it was suggested that a
zero field Hall conductivity could be induced by appropriate strain\cite{ran}.
Interesting transport phenomena are predicted, related to the
Adler-Bell-Jackiw anomaly of Weyl fermions, in parallel electric and
magnetic fields\cite{nielsen}.  In general, taking into account the interplay of
various types of disorder, scattering, and interactions in the WSM
makes modeling transport challenging.

A nice point of consistency of the proposed AIAO magnetic
order is that it is characterized on Landau symmetry grounds by a
simple Ising order parameter, for which a continuous thermal
transition is allowed. As remarked earlier, the thermal
MITs indeed appear continuous in the experiments.
Adding finite temperature to the simple mean-field Hubbard model
calculation further corroborates the expectation from Landau
reasoning, showing that the magnetic transition is continuous, and its
critical temperature grows with $U$, following the behavior of the
charge excitation gap\cite{krempa_go13}.  It may be interesting to study the
corresponding thermal and quantum phase transitions in the future, for which the
mean-field analysis may be inadequate.

\subsubsection{The role of many-body effects}  
 
In the preceding, we considered correlations only at the mean field level.
This is expected on general grounds to be qualitatively correct for
describing many phenomena within the phases so obtained.  For example,
in fully gapped states such as the TI or gapped AIAO phase, the gap
cannot be broken by any small perturbation.  The WSM is
also stable to interactions, which are irrelevant (marginally for
long-range Coulomb) in the renormalization group sense, though they
may have important transport consequences.  A cellular dynamical mean
field theory\cite{Kotliar2006,cluster-rev} (CDMFT) study bears out the
robustness of the mean field treatment, though it shows that
correlations may yet induce new phases: an axion insulator state
appears in the CDMFT analysis\cite{ara-pyro1,ara-pyro2} of \req{Hgen} (plus Hubbard $U$) though
it does not arise at the Hartree-Fock level.  For these calculations, a
convenient formulation of the $\mathbb Z_2$ topological invariant in terms of the
{\em interacting} electron Green's function\cite{Wang2012} was employed.

Other qualitative effects of correlations exist.  Most obviously, it is
likely that at least some of the quantum phase transitions indicated in
Figure~\ref{fig:hf} are not fully captured by mean-field theory.
Excitations and collective modes, which may be measured in the future
by resonant inelastic x-ray scattering (RIXS) or other methods, typically mix elements of localized and
itinerant systems in the intermediate correlation regime.  Of course, neither
CDMFT nor simpler Hartree-Fock theory are capable of capturing the
more subtle quantum correlations in other proposed phases such as the
topological Mott insulator, but we do not know that such phases
can be found in the iridates.  

\subsubsection{Interactions with rare earth moments}
\label{sec:fd-exchange} 

We now briefly turn our attention to the interactions between the
$R$-site $f$-electrons and the Ir $d$-electrons.  Depending on the
electronic configuration of the ion $R^{3+}$, the $f$-electrons can
carry a net magnetic moment being in a Kramers doublet ($R =$ Nd, Sm,
Gd, Dy, Yb) or a non-Kramers one (Pr, Tb, Ho).  Generally, if the Ir
ions magnetically order, the rare earth spins must experience an
exchange field, and also order at low temperature.  Since $4f$ spins are
rather highly localized, the exchange coupling between rare
earth and Ir spins is expected, however, to be relatively weak, so the
back-reaction on the Ir dynamics would be expected to be small.
This is consistent with experiments\cite{disseler-magOrder-y-yb} on Yb-227 and
Y-227 which show spontaneous local fields in $\mu$SR at
comparable temperatures ($T_M$ = 130 K, 150 K, respectively), though Yb$^{3+}$
carries a net moment while Y$^{3+}$ does not.  This is presumably Ir magnetism.  Signs of Yb spin ordering or freezing 
appear only below $T^* \approx$ 20 K.\cite{disseler-magOrder-y-yb}

Nevertheless, the $R$ spins can have a substantial influence on low energy
properties, and their magnetism may be non-trivial.  The most notable
example is Pr$_2$Ir$_2$O$_7$, which remains metallic to low temperature,
and shows no clear magnetic ordering transition\cite{nakatsuji-pr-06}.  Instead, novel
freezing and non-Fermi liquid phenomena appear at low temperature,
apparently driven by Pr magnetism\cite{nakatsuji-pr-06}.  Most strikingly, a zero field  
anomalous Hall effect has been observed below 1.5 K despite the absence
of magnetic order or freezing for $T>0.3$ K, leading to the proposal that
this is a realization of a chiral spin liquid\cite{nakatsuji-pr,nakatsuji07}.  Spin-ice type physics
for the Pr moments, {\it i.e.}\ predominantly 2-in/2-out configurations on each
tetrahedron, has also been inferred.  Several theoretical works\cite{onoda10,onoda11,flint13,sungbin} have 
suggested origins for these phenomena, including ordinary and exotic
RKKY interactions, quadrupolar ordering, etc. As of this writing, the physics of Pr$_2$Ir$_2$O$_7$ remains
unresolved. Overall, much less theoretical attention has been paid to the
physics of other rare earths $R$, but one work suggests their coupling to Ir
may help to stabilize the WSM and axion insulator phases.\cite{chen12}
 
\subsubsection{Issues and Outlook}

This section has outlined the theoretical and experimental work on the
pyrochlore iridates.  Both support a picture of the high-temperature state as
a paramagnetic semi-metal, possibly a zero gap semiconductor.  The
MIT or crossover occurs simultaneously with the
onset of commensurate magnetic order, most likely of AIAO type.  While
the transport phenomenology in these materials is very complex, and
clearly requires more theoretical modeling, the rough behavior of the
resistivity is in accord with static and dynamical mean-field-theory results:
the on-set of the AIAO order would generally open up a charge gap in an otherwise 
metallic or semi-metallic paramagnetic state and the MIT temperature would become smaller as the relative correlation 
strength $U/t$ gets weaker\cite{krempa_go13}.   
Understanding low temperature transport remains a challenge.

A direct determination of the magnetic structure of each member of the
series is desirable, to confirm or disprove the AIAO for all or some
compounds.  It has been argued that the AIAO order observed at the
Nd-sites in Nd-227 provides indirect evidence for the AIAO at the Ir
sites, as the ordering at Nd sites may be caused by an effective field
from the Ir's\cite{takagi-nd}.  
Recent resonant X-ray diffraction measurements on Eu-227\cite{nakatsuji-RXD}  
find $\b q=0$ magnetism with no indication of any lowering of lattice symmetry, 
prompting the authors of Ref.~\onlinecite{nakatsuji-RXD} to suggest that it must be the AIAO AF.

We have focused above primarily on possible electronic phases, but it is
also important to understand various experimental observables.  In
particular, the excitation spectrum of the insulating and conducting
states should be determined.  An understanding of
the effects of disorder and interactions on Weyl fermions is demanded to
confront spectroscopic studies.  As an example, the optical conductivities of Nd-227 and Rh-doped Nd-227 were recently measured\cite{tokura12}. The optical conductivity of
Nd-227 has a striking charge gap while Rh-doped Nd-227 shows low-energy
spectral weight that may be interpreted as an evidence for a WSM\cite{tokura12}
(though in our opinion many other explanations are possible for the latter).
There should also be collective magnetic excitations (magnons) and
possibly excitons, which may be studied by resonant inelastic X-ray
scattering (RIXS)\cite{takagi214_2012,yjkim214_2012,bjkim214_2012,bjkim327_2012,takagi327_2012,takagi327_2013} 
or other techniques.  Theoretical calculations of both
optics\cite{krempa_go13,ara-pyro2} and magnetic excitations\cite{eric_lee13}
already exist for the phenomenological
Hubbard model given in this review.  Direct examination of surface
states by photoemission or tunneling is challenging but of great interest.

\section{Strong Mott Regime} 
\label{strongU}   

In this section, we consider ``strong'' Mott insulators, by which we
mean materials for which it is sufficient to regard electrons as being
localized effectively to single atoms, and a description in terms of local spin and
orbital degrees of freedom applies.  This requires {\em a
  priori} that the charge gap is large compared to the energy of spin
and orbital excitations.    In reality, spin-orbital exchange
Hamiltonians may also provide an intuitive description of some phenomena
even in intermediate correlation situations, where charge fluctuations
are significant.  We will not comment further on this, and proceed
with the strong Mott discussion.  

The role of strong SOC in the strong Mott regime is clearly very
different than that in more itinerant systems, since if electrons can be regarded as wholly localized, then
considerations of {\em band} topology do not apply.  In the strong
Mott regime, instead, the interesting physics of SOC arises from the
unique way in which it -- wholly or partially -- resolves {\em orbital
degeneracy}.    Orbital degeneracy is germane to correlated materials,
in which the environment of transition metal atoms often has at least
approximate octahedral or tetrahedral symmetry, resulting in
degenerate doublets or triplets of orbitals.  When these shells are
neither half-filled nor full, orbital degeneracy arises.  {\sl In
  principle}, this leads, even without SOC, to interesting physics, in
which the orbital degree of freedom behaves as an additional
``pseudo-spin'' quantum variable.  The combined exchange of spin and
pseudo-spin is given then by {\em Kugel-Khomskii
  models}\cite{Kugel82}.  In practice, the ``quantumness'' of the
orbitals is usually compromised by the Jahn-Teller effect, in which 
lattice distortions spontaneously arise to split the orbital
degeneracy \cite{JT}.  Even at temperatures above this orbital ordering one, the
associated phonon modes couple strongly to the orbital pseudo-spin,
damping and decohering it.  Consequently, a sadly mundane reduction of
the beautiful Kugel-Khomskii Hamiltonian, by simply replacing the
orbital pseudo-spin by its classical expectation value, seems to be
adequate for the description of many orbitally degenerate transition
metal systems, such as perovskite manganites.  Even in the rare
situations in which something more subtle is suspected, the mixing of orbital and lattice modes makes it practically difficult to clearly distinguish collective
orbital excitations in experiments. 

SOC offers a more interesting way of resolving orbital degeneracy,
trading the Jahn-Teller effect for {\em entanglement} of spin and
orbital degrees of freedom.  It may entirely suppress the orbital
degeneracy, leaving behind a pure Kramers doublet, as in the case of
the $J_{\rm eff}=1/2$ states of Ir$^{4+}$ ions, or even selecting a
spin-orbital singlet state for some non-Kramers ions, as in the case
of tetrahedral Fe$^{2+}$\cite{chen2008sos}.  In other cases the
degeneracy lifting may be partial, leading to effectively larger
spin-orbital pseudo-spins.  In either the Kramers or the latter case,
the interactions amongst the remaining highly entangled states are
strongly affected by this entanglement and the underlying spin-orbital
exchange.  Consequently, not only do such systems (at least largely) avoid the
Jahn-Teller effect, but they enjoy exotic exchange interactions which
foster unusual ground states.  In the remainder of this section, we
will discuss in particular the possibilities of {\em quantum spin
  liquid} and {\em multipolar ordered} phases, as well as
unconventional magnetically ordered states.  We illustrate these
possibilities through two material examples, the honeycomb iridates
and the double perovskites.

\subsection{Full degeneracy lifting and honeycomb iridates}
\label{sec:honeyc-irid-kita}

As discussed already in Sec.~\ref{pyro}, for an octahedrally coordinated
Ir$^{4+}$ ion with 5$d^5$ configuration, SOC completely removes
orbital degeneracy, resulting in a maximally quantum effective
spin-1/2 Hamiltonian, representing the $J_{\rm eff} = 1/2$ states.
One can say that the orbital degeneracy is fully lifted, the
remaining degeneracy being guaranteed by Kramers theorem.  The
hexagonal iridates Na$_2$IrO$_3$ and Li$_2$IrO$_3$, which realize a
layered structure consisting of a honeycomb lattice of Ir$^{4+}$ ions,
provide a concrete example of this case.  Both compounds appear to be
in the strong Mott regime.   As shown by Jackeli and
Khaliullin\cite{jackeli2009mott}, in the ideal limit, the edge sharing
octahedral structure and the structure of the entangled $J_{\rm
  eff}=1/2$ orbitals leads to a cancellation of the usually dominant
antiferromagnetic oxygen-mediated exchange interactions.  A
sub-dominant term is generated by Hund's coupling, which takes the
form of highly anisotropic exchange:
\begin{equation}
H_K = - K \sum_{\alpha=x,y,z} \sum_{\langle ij\rangle \in \alpha} S^{\alpha}_i S^{\alpha}_j \ , \label{eq:1}
\end{equation}
where ${\bf S}_i$ are the effective spin-1/2 operators, and 
$\alpha=x,y,z$ labels both spin components and the three orientations
of links on the honeycomb lattice.   This peculiar and highly frustrated
Hamiltonian is, remarkably, nearly the only example of an exactly
soluble model for a quantum spin liquid state!  As shown in an
ingenious and tour de force paper by Alexei
Kitaev\cite{Kitaev20062}, it describes a state with no magnetic order and elementary excitations which
are charge-neutral ``spin"-carrying Majorana fermions that are their
own anti-particles.  It is astonishing that the Kitaev exchange form
of Eq.~\eqref{eq:1}, which is very unnatural for conventional magnetic
systems, arises organically from the geometry and
entanglement in the strong SOC limit.  

The experimental situation is more complex.  In
this\cite{PhysRevLett.105.027204,PhysRevLett.110.097204} and
other\cite{Chen08} contexts, it has been argued that an isotropic
Heisenberg interaction is generated from the direct overlap of 5$d$
orbitals.  The resulting Heisenberg-Kitaev model has been studied by
several methods, which demonstrate that with increasing Heisenberg
coupling the ground state undergoes successive transitions from the
Kitaev spin liquid to a ``stripy'' four-sublattice magnetically
ordered state, to the usual two sublattice one. Recent neutron
scattering experiments and other studies show that the ground state of
Na$_2$IrO$_3$ is, however, none of these states, but instead displays
a different collinear magnetic order, the so-called zig-zag state with
a four-sublattice structure\cite{yjkim213_2011,PhysRevLett.108.127204}.  This has generated a number of new
theoretical proposals to explain the apparent departure from the
Heisenberg-Kitaev model\cite{mazin_2012,trebst_2011,kimchi_2011,
subhro213_2012,ronny13}.  Possibly important ingredients are other
symmetry-allowed interactions on nearest-neighbor bonds, which might
emerge from trigonal crystal fields or other mechanisms, and exchange
with second and third neighbor sites.  However, it has been suggested
that the Kitaev interaction might be much larger in Li$_2$IrO$_3$ and
the system may be closer to the quantum spin liquid
phase\cite{gegenwart213_2012}.  It is perhaps worthwhile to comment on
the role of frustration.  Often in antiferromagnets, SOC, for example via the
Dzyaloshinskii-Moriya interaction, is thought to remove accidental
degeneracy and favor order.  The Kitaev model is a counterexample,
showing that in some cases strong SOC can suppress ordering.  However,
one should be aware of both possibilities.

\subsection{Partial degeneracy lifting and ordered double perovskites}
\label{doubleP}
 
In other situations, SOC may reduce but not eliminate fully the
orbital degeneracy.  Such partial degeneracy removal still has dramatic
effects upon the physics.  As an example, we consider a family of
compounds with one or two electrons in the 4$d$ or 5$d$ shell.  These are
thus moderate to strongly spin-orbit coupled analogs of the very well
studied $3d$ titanates with Ti$^{3+}$ and vanadates with V$^{3+}$ or
V$^{4+}$ states.  The latter materials constitute classic families undergoing Mott transitions\cite{mott}.  
In octahedral coordination and
for vanishing SOC, these valence states possess a three-fold orbital
degeneracy, and consequently are Jahn-Teller active in the solid and
tend to exhibit complex orbitally ordered states.  

When SOC is dominant, we can understand the partial degeneracy lifting
as follows.  The $t_{2g}$ orbitals split just as in iridates, but with
the 1 or 2 electrons occupying now the $J_{\rm eff}=3/2$ quadruplet.
In the $d^1$ case, this simply results in a $J_{\rm eff}=3/2$ local
effective spin.  In the case of the $d^2$ electron configuration, the
description based purely on single-particle states no longer holds,
but the result is similar.  Following Hund's rules applied to the
$d^2$ configuration, one finds a total spin $S=1$ and an effective
orbital angular momentum (three-fold degeneracy) $L_{\rm eff}=1$.  In
the end $S$ and $L$ align, so that an effective $J_{\rm eff}=2$ spin
results when SOC is included\cite{Goodenough68,chen11}.  We see that
in both cases, the ionic degeneracy ($=4,5$ for $d^{1}$ and $d^{2}$,
respectively) is reduced from that without SOC ($=6,9$, respectively),
but is still larger than what is required by Kramers theorem alone
($=2,1$).  

What physics can we expect from the effective
spins?  ``Large'' spins such as $S=3/2,2$ are often thought to behave
relatively classically.  This is based on the standard spin wave
expansion for Heisenberg-type models, which indeed can be cast as an
expansion in $1/S$.  However, the classicality of such spins in fact
depends critically on the nature of their interactions.  Many examples
of non-classical behavior for larger spins can be found in the
theoretical literature, notably spin-nematic or quadrupolar ordered
phases of spin one systems with biquadratic exchange\cite{Andreas06},
Haldane type states for $S\geq 1$ AKLT models\cite{AKLT}, and quantum spin liquid states
of large $N$ SU(N) and Sp(N) antiferromagnets\cite{sachdev_review}.  All these cases can be
cast in the form of models in which spins with $S>1/2$ interact with
{\em higher order exchange} interactions involving multiple spin
operators on each site (but typically with spins still interacting
pairwise).  One can describe such interactions as consisting of
coupling between {\em multipole} moments (beyond dipoles) of the
spins.  Heuristically, multipolar interactions enhance quantum
fluctuations because they connect directly states with very different
$S^z$ quantum numbers, allowing the wavefunction to easily delocalize
in the spin space.  By contrast, the usual bilinear Heisenberg
interactions induce only single spin flips, and consequently
tend to localize the spin wavefunction near some classical extremum.

Unfortunately, multipolar interactions are very weak in conventional
systems with weak SOC.  However, this is not true for the effective
spins in the strong SOC limit.  This is because spin exchange is {\em
  always} dependent on the orbital state, and in the strong SOC limit
spins and orbitals are highly entangled.  This entanglement transfers
the orbital dependence of the exchange to the effective spin,
generating multipolar exchange, as explained in
Sec.~\ref{sec:multipolar-exchange}.  Thus for spin-orbitally entangled
effective spins, {\em strong multipolar exchange is generic}. We
illustrate this in the following subsections for one class of
materials.

\subsubsection{Double perovskites}
\label{sec:double-perovskites}

\begin{figure}[b]
\subfigure[]{\label{fig:fig0} \includegraphics[width=10cm]{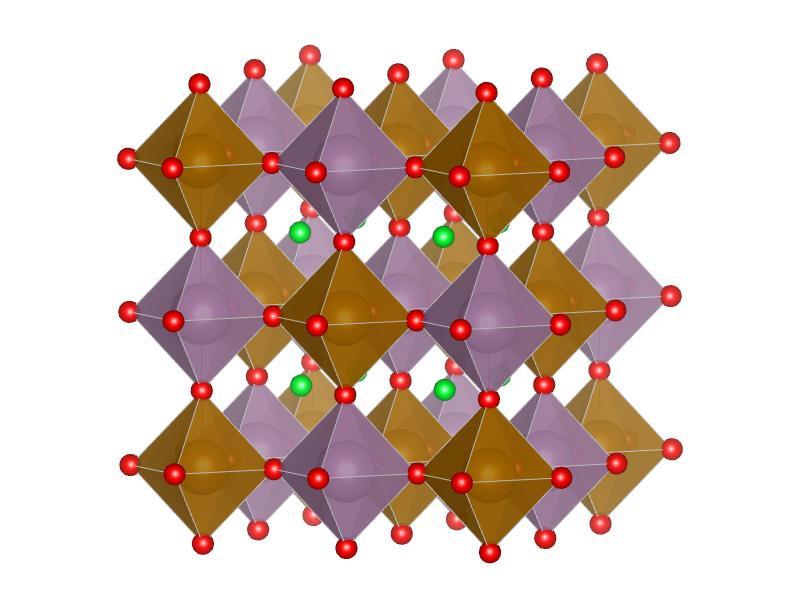}}
\subfigure[]{\label{fig:orb} \includegraphics[width=7.1cm]{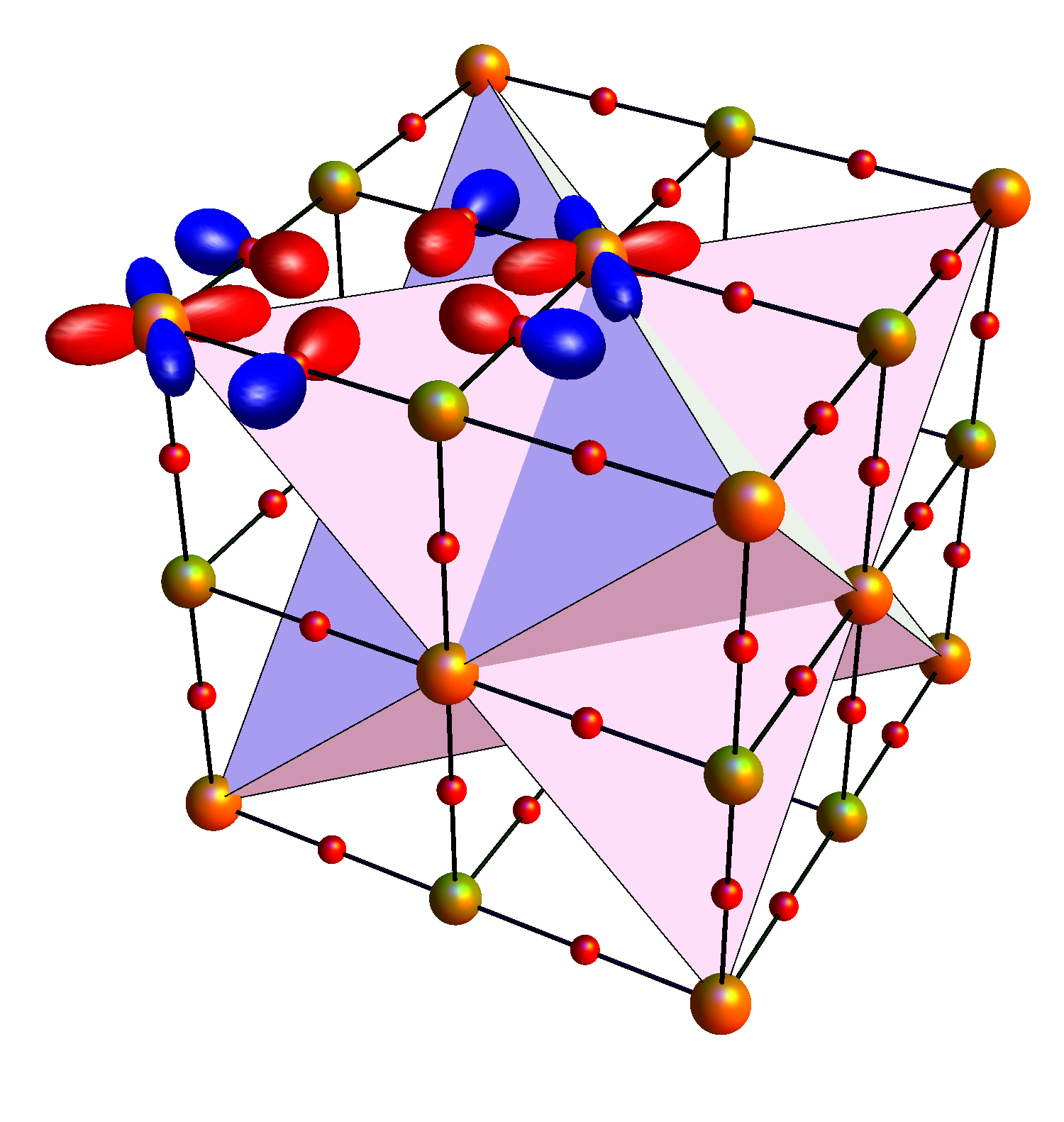}}
\caption{a) The crystal structure of ordered double perovskite,
  $A_2BB'$O$_6$. The green spheres, purple octahedra, and dark
  yellow octahedra represent $A$-site ions, $B$O$_6$ octahedra, and
  $B'$O$_6$ octahedra, respectively.  b) The same structure,
  showing the representation of the geometrically frustrated fcc
  lattice of $B$ sites (orange spheres) as edge sharing tetrahedra.  Two
  $d_{xy}$ orbitals on nearest-neighbor $B$ sites are shown with the
  intermediate $p_x$,$p_y$ orbitals involved in their exchange path.
  In this picture $B'$ sites are golden, and O sites are red, and $A$ sites
are not shown.  }
\label{fig0}
\end{figure}

For numerous examples of partially quenched degeneracy of the above
type, we turn to the double perovskite materials, $A_2BB'$O$_6$,
whose structure is shown in \rfig{fig0}.  It can be derived from
the simpler cubic perovskite structure, $AB$O$_3$, but replacing the
single $B$ atom by alternating $B$ and $B'$ atoms on two rock-salt
sublattices.  There are many such compounds with non-magnetic $B$ sites
and $B'$ sites occupied by the 4$d$ and/or 5$d$ transition metal
elements with $d^{1}$ or $d^{2}$ configuration.  Because of the
difference in the valence charges and ionic radius between $B$ and $B'$
ions, the magnetic $B'$ ions form an fcc lattice structure with very
little intersite disorder, and the lattice constant twice that of the
original cubic case.  As shown in Table \ref{tab2}, the magnetic ions
$B'$ (Re$^{6+}$, Os$^{7+}$, Mo$^{5+}$ for $d^1$ or Re$^{5+}$,
Os$^{6+}$ for $d^2$) have one or two electrons on the $4d$ or $5d$
shell\cite{Cussen06,Vries10,Aharen10,Carlo11,Qu13,Vries13,Aharen102,Wiebe03,Wiebe02,yamamura06,Stitzer02,Erickson07,Steele11,Aharen09}. 
Like Iridium, such heavy
elements and high orbital shells naturally incorporate strong SOC.  Moreover, the large separation of the metal ions
in this structure suppresses electron hopping and promotes Mott
insulating behavior.

Immediate evidence for orbital degeneracy lifting can be found in several of
these materials.  One consequence of the strong SOC limit is a
reduction of the effective moment.  For the $J_{\rm eff}=3/2$ state,
in fact, in an ideal isolated ion, the orbital and spin contributions
cancel, and the moment {\em vanishes}, \emph{i.e.}\ the g-factor is zero!  Due
to mixing with neighboring oxygen ions, the magnetic moment is
generally non-zero in the solid, but a small g-factor is still
expected.  The situation for $J_{\rm eff}=2$ is similar, but less
dramatic: the magnetic moment is reduced to ${\b M} = {\b S}_{\rm
  eff}/2$, \emph{i.e.}\ $g=1/2$.  Indeed, the magnetic moments of most
compounds in Table \ref{tab2} are smaller than the values expected
from a spin-only contribution, consistent with the effective local
moment renormalized by the strong SOC\cite{chen10,chen11,Lee07,Xiang07}.  

Another indicator of strong SOC is the magnetic entropy, which can
give direct evidence of the quadruplet or quintuplet structure.  In
Ba$_2$YMoO$_6$, an entropy of $R \ln 4$ (here $R$ is the gas constant) was indeed estimated from
integration of the magnetic specific heat\cite{Vries10}, consistent with the $J_{\rm eff}=3/2$ state.  

\begin{table}[h]
  \caption{
    A list of representative ordered double perovskites in which the magnetic ions have $d^1$ 
    or $d^2$ electron configuration. 
    Variations in the Curie-Weiss temperature $\Theta_{CW}$ and effective
    magnetic moment $\mu_{\text{eff}}$ may originate from the experimental fitting 
    of data at different temperature ranges. In the Table, PM = paramagnetic, 
    AFM = antiferromagnetic, and FM = ferromagnetic phase. $T_c$ is the ordering 
    temperature, and $T_G$ is the glass transition temperature. 
  }
\begin{center}
\begin{tabular}{ccccccc}
Compound & $B'$   & electron config. & $\Theta_{CW}$ (K) & $\mu_{\text{eff}}$ ($\mu_B$) & 
magnetic transition & Refs 
\\
\hline\hline
\\
Ba$_2$YMoO$_6$ &  Mo$^{5+}$ &  4$d^1$ & $-91$ $\sim -219$ & $1.34\sim 1.72$ & PM down to 2K & 
  \onlinecite{Cussen06,Vries10,Aharen10,Carlo11,Qu13,Vries13}
\\
\\
Sr$_2$MgReO$_6$ &Re$^{6+}$ & 5$d^1$ & $-426$ & 1.72 & spin glass, $T_G \sim 50$K &
\onlinecite{Wiebe03}
\\
\\
Ba$_2$NaOsO$_6$ & Os$^{7+}$ & 5$d^1$ & $\sim -10$ &  $\sim0.6$ & FM $T_c =6.8$K& 
\onlinecite{Erickson07}
\\
\\
\hline\hline
\\
Ba$_2$CaOsO$_6$ & Os$^{6+}$ &  5$d^2$  & $ -157$     & $ 1.61$        & AF $T_c = 51$K & 
\onlinecite{yamamura06}
\\
\\
La$_2$LiReO$_6$  &  Re$^{5+}$ & 5$d^2$ & $-204 $    & $1.97$         &  PM down to 2K &
\onlinecite{Aharen102}
\\
\\
\hline\hline
\end{tabular}
\end{center}
\label{tab2}
\end{table}
\subsubsection{Multipolar exchange}
\label{sec:multipolar-exchange}

We now explain how multipolar exchange arises from spin-orbital
entanglement, a known phenomena in $f$-electron
systems.\cite{RevModPhys.81.807} The basic idea is to consider the
Kugel-Khomskii type exchange that arises when all orbitals are
included, and then to {\em project} that exchange to the effective
spins that form in the strong SOC limit.  In general, several
different exchange processes contribute to the appropriate
Kugel-Khomskii model for double perovskites.  For simplicity, we will
just illustrate one in detail here, and refer the reader to
Refs.~\onlinecite{chen11,chen10} for more details.  We consider the $d^1$
case, and focus on the nominally antiferromagnetic nearest-neighbor
process via intermediate oxygens, which would be expected to dominate.
These processes are strongly restricted by orbital degrees of freedom.
As is illustrated in \rfig{orb}, in $xy$ planes, only
electrons residing on $d_{xy}$ orbitals can virtually transfer to
neighboring sites via $p_x$ and $p_y$ orbitals of the intermediate
oxygens (the same process may be understood as direct exchange between
molecular orbitals consisting of transition metal $d_{xy}$ and neighboring
oxygen $p$ levels).  Therefore, the antiferromagnetic exchange
interaction can be written as ${\mathcal H}_{\text{ex}} = {\mathcal
  H}^{xy}_{\text{ex}} + {\mathcal H}^{yz}_{ \text{ex} } + {\mathcal
  H}^{xz}_{\text{ex}}$
with 
\begin{equation}
 {\mathcal H}^{xy}_{\text{ex}} = J \sum_{\langle ij \rangle \in xy \text{ plane}} 
 \Big(   {\b S}_{i,xy} \cdot {\b S}_{j,xy}  - \frac{1}{4} n_{i,xy} n_{j,xy}   \Big),\label{eq:3}
\end{equation}
where the sum is over nearest neighbor sites in the $xy$ planes, and the corresponding terms 
in $yz$ and $xz$ planes are obtained by a cubic permutation. Here the operators  
${\b S}_{i,xy}$ and $n_{i,xy}$ denote the spin residing on the $d_{xy}$ orbital and $d_{xy}$-orbital occupation
number at site $i$, respectively.  

Without SOC, and as written, the above interaction appears relatively
conventional and in particular is bilinear in ${\b S}_{i,xy}$.  However,
this can be rewritten by explicitly representing the orbital degree of
freedom via the effective $L=1$ angular momentum ${\b L}$ describing
the $t_{2g}$ degree of freedom.  One has ${\b S}_{i,xy} = \b S_i^{\rm tot} [1  
- (L_i^z)^2]$ and $n_{i,xy} = 1 -(L_i^z)^2$, where $\b S_i^{\rm tot}={\b S}_{i,xy}+{\b S}_{i,xz}+{\b S}_{i,yz}$ is the
total true spin on site $i$.  With these substitutions, we see that
up to 3 spin or pseudo-spin
operators are multiplied on each site $i$ or $j$. 
Now in the strong SOC
limit, \emph{i.e.}\ $J \ll \lambda$, this should be projected onto the 
$J_{\rm eff}=3/2$ effective spin.   That is, we should replace
${\mathcal H} \rightarrow \tilde{\mathcal H}$, in which operators on
each site have been replaced by their projections,  $\tilde{\mathcal O}
= {\mathcal P}_{\frac{3}{2}} {\mathcal O} {\mathcal P}_{\frac{3}{2}}$,
where ${\mathcal P}_{\frac{3}{2}}$ is the projection operator onto the $J_{\rm eff}=3/2$ multiplet. 
Some algebra shows that these projections are quite non-trivial:
\begin{equation}
\tilde{S}^{\alpha}_{i,xy} = \frac{1 + 2 \delta_{\alpha, z}}{4} S_i^{\alpha} - \frac{1}{3} S_i^z S^{\alpha}_i S_i^z  \ \ \ (\alpha = x, y, z),  \ \ \ \ 
\tilde{n}_{i,xy} = \frac{3}{4} -\frac{1}{3} (S_i^z)^2,\label{eq:4}
\end{equation}
where ${\b S}_i$ is the final projected $J_{\rm eff}=3/2$ effective
spin.  Spin and occupation number operators for other orbitals can be readily generated by a cubic
permutation. 

The quadratic and cubic products of $S_i^\mu$ represent
components of the quadrupolar and octupolar tensors, respectively.  We
see that the innocuous-looking Hamiltonian in Eq.~\eqref{eq:3} is
transformed, after the strong SOC projection using ${\mathcal O}
\rightarrow \tilde{\mathcal O}$ via Eq.~\eqref{eq:4}, into a highly
non-trivial interaction with octupolar and quadrupolar
couplings of the same order as ordinary bilinear exchange.   This mechanism
generating higher-order spin exchange is quite generic, and applies
equally to all the interactions in a typical strong SOC situation.  It
gives access to a variety of exotic physics of multipolar systems\cite{Santini09,Shiina98}.

For the $d^1$ double perovskites, two natural additional exchange
channels were identified between nearest-neighbors in Ref.~\onlinecite{chen10}:
a ferromagnetic exchange $J'$ between orthogonal orbitals, and an
electrostatic quadrupole interaction $V$.  Details of these interactions
can be found in Ref.~\onlinecite{chen10}.   Similar considerations apply to the
$d^2$ case, as described in Ref.~\onlinecite{chen11}.  All the interactions in
both cases become multipolar in character after the strong SOC
projection.  

\begin{figure}[t]
\includegraphics[width=11cm]{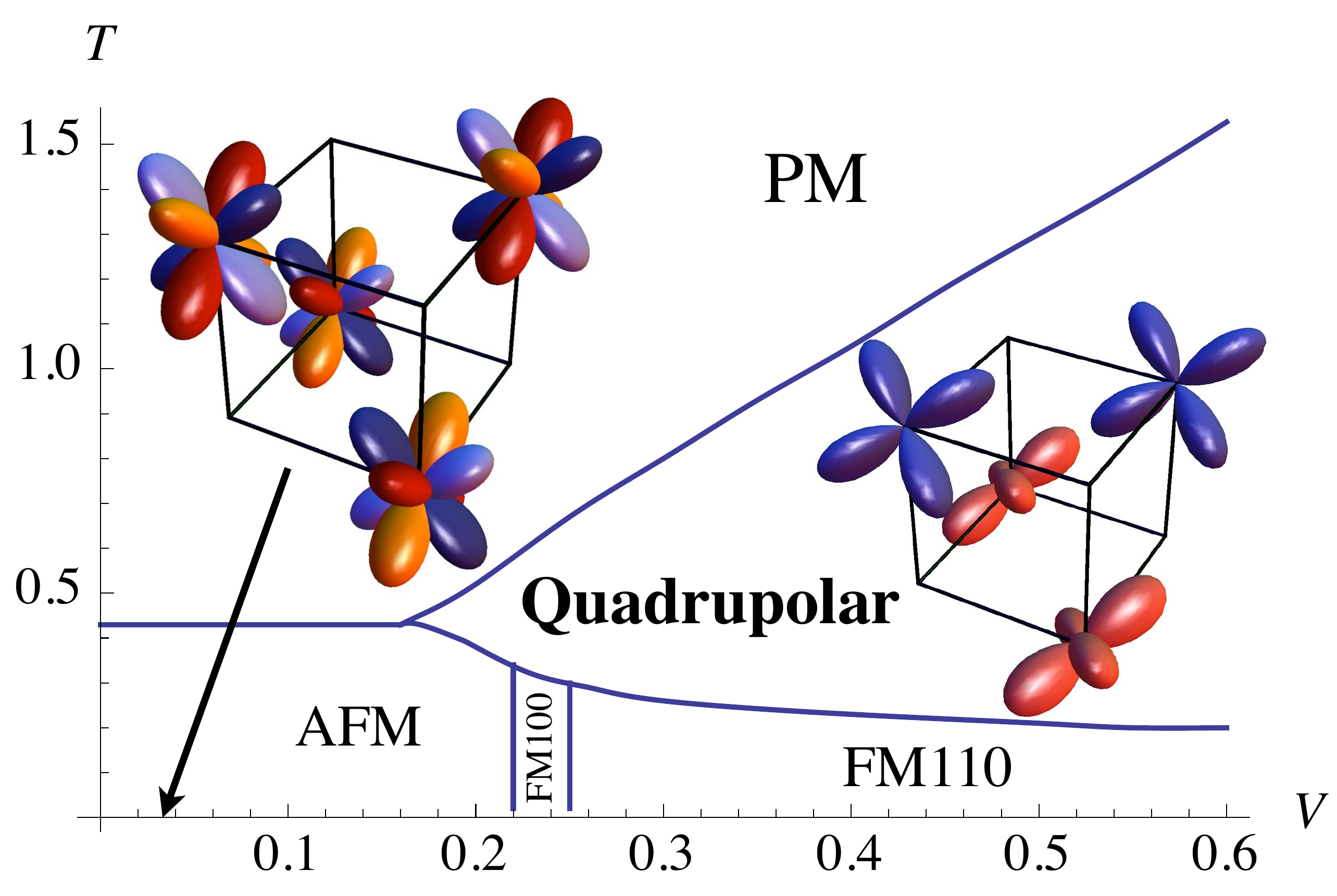}
\caption{A cut of the mean field phase diagram for $d^1$ double perovskites
  at fixed $J'=0.2J$, and $J=1$, as a function of temperature $T$ and
  electric quadrupole interaction $V$.   The antiferromagnetic (AFM)
  state is illustrated at $T=0$ by an image of the orbital
  wavefunctions for $S^x = + 1/2$ (with positive/negative regions
  colored blue/light blue) and for the $S^x=-1/2$ (with positive/negative
  regions colored red/orange).   In the quadrupolar state, the charge
  density is shown.  The FM110 and FM100 are ferromagnetic states with
  net magnetization along the [110] and [100] axes (and spin-orbital
  entanglement which is difficult to illustrate).  The curves are
  obtained from calculations in Ref.~\onlinecite{chen10}.}
\label{fig3}
\end{figure}

\subsubsection{Mean field theory}

With the exotic interaction Hamiltonians discussed above, unusual
phases are expected even in a mean field treatment.  We explore this
here, and subsequently consider the possibility of phases beyond mean
field theory.  The Weiss mean field treatment consists, as usual of
decoupling the interactions between sites, to generate self-consistent
single site problems.  Due to the multipolar interactions, each single
site Hamiltonian contains not only an effective Zeeman field, but also
effective quadrupolar and octupolar anisotropies.  The latter define
additional multipolar order parameters in the mean field theory.  A
full analysis is given in Ref.~\onlinecite{chen10} for the $d^1$ case and
Ref.~\onlinecite{chen11} for the $d^2$ case.  We summarize only the former
as an example.  

A representative cut through the three-dimensional phase diagram is depicted in Figure \ref{fig3}.
At zero temperature three phases appear,
all with nominally ``conventional'' ferromagnetic or antiferromagnet
dipolar magnetic order.    However, there are unconventional aspects
revealed upon closer inspection.  

The simplest state is the antiferromagnetic phase, appearing for small
$J'/J$ and $V/J$.  It has a conventional two-sublattice structure,
with states on either sublattice related by time reversal.  However,
although it displays local static dipole moments, their magnitude is
small, and in fact {\em vanishes} as temperature $T\rightarrow 0$.
This is particularly surprising in a mean field theory (for a
conventional Heisenberg system mean field theory gives a full
moment).  In fact, accompanying the dipole moment is a large {\it
  staggered octupole moment}, which competes with and suppresses
substantially the dipolar order. 

With larger $J'/J$ and $V/J$, the system develops ferromagnetic phases, 
FM110 and FM100 with net ferromagnetic moments along [110] and [100] directions, respectively. 
These two non-uniform ferromagnets are rather unconventional as they
actually have a two sublattice structure, with partial cancellation of
non-parallel magnetic moments in the ferromagnetic magnetization.  In
fact, the two sublattice structure is a manifestation of staggered
quadrupolar order, and it is this quadrupolar ordering which
predominantly drives the formation of these two phases.  The magnetism
develops atop it.  Since orbital polarization is distinct on the two
sublattices, they cannot be time-reversal conjugates, and consequently
when magnetism onsets, a net ferromagnetic moment results.  

The driving role of the quadrupolar order can be seen from the $T>0$
phase diagram.  Over a wide range of intermediate temperature, the
ferromagnetic order is destroyed, the FM region (and part of the AFM
one) being replaced by a purely quadrupolar ordered phase.  In the
quadrupolar phase, time-reversal symmetry is unbroken, which is
sufficient to require the dipolar and octupolar order parameters to
vanish.  A standard classification scheme for quadrupolar states is to
examine the eigenvalues of the traceless quadrupolar tensor
$Q_i^{\mu\nu} = \langle S_i^{\mu} S_i^{\nu}\rangle-\frac{1}{3} S(S+1)
\delta^{\mu\nu} $, where the eigenvalues must sum to zero (here
$S=3/2$ for $d^1$).  The quadrupolar phase with only one independent
eigenvalue, {\it i.e.}\  ${\rm eigenvalues}(Q) =\{ q,q,-2q\}$, is
called the {\sl uniaxial nematic phase}, and corresponds to the
situation, where one principal axis is distinguished from the other
two that remain equivalent.  This type of ``spin nematic'' has been
studied theoretically in $S=1$ Heisenberg models with strong
biquadratic interactions,\cite{Andreas06} though it is hard to achieve
such strong biquadratic exchange in conventional systems.  In the most
general case, there may be two independent eigenvalues, {\it i.e.}\
${\rm eigenvalues}(Q) = \{q_1,q_2,-q_1-q_2\}$, with $q_1\neq q_2$.
This is called a {\sl biaxial nematic phase}, where all three
principal axes are non-equivalent. The quadrupolar state obtained here
is such a biaxial nematic phase. 

Such a quadrupolar phase also exists in the phase diagram for the
$d^2$ double perovskites\cite{chen11}, but in that case appears even
at $T=0$.  This difference between the $d^1$ and $d^2$ cases occurs
because within Weiss mean field theory, a Kramers ion such as $d^1$
must always break time reversal symmetry at $T=0$ to avoid ground
state degeneracy.

\subsubsection{Beyond mean-field theory}
\label{sec4sub4}

As remarked earlier, multipolar interactions tend to destabilize
conventional, magnetically ordered semiclassical ground states.
Roughly speaking, this is because they contain many more ``spin flip''
terms analogous to the $S_i^+ S_j^-$ couplings which are part of the
usual Heisenberg two spin interactions.  Quantum disordered ground
states can be established rigorously for AKLT models\cite{AKLT}, which have
specially tuned SU(2) invariant Hamiltonians which can be written
entirely in
terms of a positive semi-definite sum of projection operators.  They
can also be found in a controlled manner for large $N$ models with
enlarged SU($N$) or Sp($N$) symmetry\cite{sachdev_review}.  Neither approach can be directly
applied here, but it is clear that the multipolar Hamiltonians which
occur naturally for $d^1$ and $d^2$ systems are somewhat intermediate
between conventional spin models and these special cases.  Thus in
frustrated geometries, quantum disordered states may well obtain.

A simple way to check for disordered states is to
gauge the magnitude of quantum fluctuations within a spin-wave
expansion (generalized for multipolar phases) about the mean field states. Indeed using a generalized Holstein-Primakoff spin wave
theory, it is shown that quantum fluctuations are strong when $J'/J$
and $V/J$ are small \cite{chen10}.  This strongly points to effects
beyond mean field, and the likelihood of a non-magnetic ground state.
By examining various limits, both valence bond solid and quantum spin
liquid states were proposed for this region of the phase diagram of
the $d^1$ systems in Ref.~\onlinecite{chen10}.   Further quantitative study of
models of this type would be highly desirable, but this is challenging
and probably requires at least significant extension of existing
numerical methods.

Other theoretical work has considered the effects of non-cubic crystal
fields, which split the quadruplet ground state to form a low energy
doublet.  This then reduces to an effective $S=1/2$ model, but with
highly anisotropic interactions with anisotropy depending on the bond
orientation.  The resulting exchange model may be strongly frustrated.
In such cases, quantum fluctuations may also support a quantum spin
liquid phase \cite{Dodds11}, which is amenable to analysis by more
conventional methods.

\subsubsection{Connections to experiments} 
\label{sec4sub5}

A detailed discussion about the applications of theory to experiment,
and on specific materials, can be found in Ref.~\onlinecite{chen10}
and Ref.~\onlinecite{chen11}.  Here we comment just on a few examples
for which strong SOC phenomenology is most compelling.  In general,
for the strong SOC limit to apply, both non-cubic crystal fields and
exchange interactions should be small compared to the spin orbit
splitting.  The latter is of order 100~meV for 4$d$ materials like
Mo$^{5+}$, and a few times larger for the 5$d$ materials.  This is quite
large compared to exchange interactions in double perovskites, but not
necessarily large compared to crystal field perturbations in non-cubic
crystals.  Thus we focus on cases where at least elastic X-ray
scattering measurements find a cubic structure.  This presumably
indicates the absence of a static Jahn-Teller effect, though the
dynamic or non-uniform effects of Jahn-Teller phonon coupling may
still be important.

We have already mentioned Ba$_2$YMoO$_6$, which remains cubic to low
temperatures and has a magnetic entropy approximately equal to $R \ln
4$, consistent with a $J_{\rm eff} = 3/2$ local moment\cite{Vries10}.  This
material avoids any apparent phase transitions down to 2~K, despite
rather strong antiferromagnetic interactions.  An energy gap is
indicated in recent NMR and inelastic neutron scattering measurements,  which
may be associated with valence bond formation\cite{Carlo11}.  An
unusual feature seen in this material, which occurs in a number of
double perovskites, is the existence of {\sl two} Curie regimes in the
magnetic susceptibility: above 100~K and below 
50~K\cite{Aharen10,Vries10,Qu13}.  This is suggestive of some single-ion anisotropy, which would 
explain the existence of two Curie regimes because it splits the 4-fold degeneracy of 
the $J_{\rm eff} = 3/2$ states but leaves a 2-fold Kramers doublet at temperatures below 50~K, 
which still gives a Curie signal.  This could perhaps be associated
with quadrupolar order, or to a lattice driven distortion.  The
macroscopic cubic symmetry appears at odds with the latter, but
a set of random local distortions with statistical cubic
symmetry might be a possibility.  More recent experiments have observed
the emergence of additional phonon mode below 130~K, consistent with a local structural 
change\cite{Qu13}. 

A second example is Ba$_2$NaOsO$_6$, for which the ground state below
6.8~K was found to be ferromagnetic with easy axis along the
[110] direction\cite{Erickson07}.  The [110] easy axis is quite
unusual, and indeed does not occur in standard Landau theory for
ferromagnets, in which weak cubic anisotropy (quartic terms) always
selects either a [100] or [111] axis.  This is a strong indication for
some correct element in the strong SOC description, which indeed
predicts [110] ferromagnetism to be dominant, due to the quadrupolar
mechanism.   Experimental confirmation would require measurement of a
structure change associated to the quadrupolar ordering, which so far
has not been observed.  However,  several other indications
are in favor of quadrupolar ordering \cite{chen10}.

\section{Concluding Remarks and Outlook}
\label{sec:concl}
The investigation of correlated electron systems with strong
spin-orbit coupling (SOC) is still in its infancy.  In this review, we
summarized and explained recent developments in theoretical and
experimental research in this rapidly growing area.  We show that SOC
tends to act with Coulomb interactions to enhance the degree of
electron correlations, producing spin-orbit-assisted Mott insulators.
In the weak to intermediate correlation regime, we described possible
electronic states with topological band structures.  When correlations
are strong enough that a local moment description applies, we
discussed how spin liquid and multipolar phases may occur due to
spin-orbital entanglement.  We explored possible applications to
pyrochlore iridates in the weak-to-intermediate correlation regime,
and honeycomb iridates and double-perovskites with 4$d$/5$d$
transition metal elements in the strong Mott regime.

Due to space limitations, many important systems were not discussed.
This includes the Ruddlesdon-Popper series of perovskite iridates\cite{SrIrO-prl,SrIrO-science,takagi214_2012,yjkim214_2012,bjkim214_2012,bjkim327_2012,takagi327_2012,takagi327_2013,hykee1,hykee2,hykee3},
Sr$_{n+1}$Ir$_{n}$O$_{3n+1}$, whose $n=\infty, 1,2$ members are some
of the most investigated materials in this class.  In particular,
Sr$_2$IrO$_4$ warrants special attention as an approximate homolog of
the parent material of some of the high-T$_c$ cuprates, La$_2$CuO$_4$.  Like 
the latter, Sr$_2$IrO$_4$ is an antiferromagnetic 
insulator\cite{SrIrO-prl,SrIrO-science,takagi214_2012,yjkim214_2012,bjkim214_2012}, with a
very large exchange constant $J \sim 700-1000$~K\cite{takagi214_2012,yjkim214_2012,bjkim214_2012}, 
somewhat smaller than but comparable to that of
the cuprates, and a charge gap $\Delta_c \sim 350-650$~meV or $4000-7500$~K, 
many times larger than $J$.  
Both Sr$_2$IrO$_4$ and Sr$_3$Ir$_2$O$_7$ have served as striking
examples of the power of modern 
RIXS, which is remarkably suited for $5d$ compounds\cite{SrIrO-science,takagi214_2012,yjkim214_2012,bjkim214_2012,bjkim327_2012,takagi327_2012,takagi327_2013}.  
RIXS has provided a direct probe of spin and orbital states and dynamics.  By
RIXS, it has been possible to test and confirm the $J_{\rm eff}=1/2$
character of the Ir$^{4+}$ state in Sr$_2$IrO$_4$ \cite{SrIrO-science} and 
to measure the full spin-wave dispersion in both 
compounds\cite{takagi214_2012,yjkim214_2012,bjkim214_2012,bjkim327_2012,takagi327_2012,takagi327_2013}.  
Such measurements reinforce the analogy of Sr$_2$IrO$_4$ to the cuprates,
and have motivated a strong push to {\em dope} this material\cite{electron_doping,Rh_doping,Mn_doping} 
in hopes of finding high-temperature superconductivity, pseudo-gap behavior,
etc, as also suggested theoretically\cite{wang-senthil}. This program
is very much underway, though at the time of this writing
superconductivity has not been found.  

In general, doping of correlated insulators with strong SOC is a very
interesting subject.  In the traditional venue, the Mott
metal-insulator {\em quantum} phase transition may be tuned either by
bandwidth control at fixed stoichiometric filling, or by filling
control, varying the carrier density, and these two types of Mott
transitions have quite distinct characters\cite{RevModPhys.70.1039}.  
In this Review we have
already remarked upon the seemingly distinct behavior of the bandwidth
controlled transition (and the thermal transition at fixed filling) in
the strong SOC regime from the usual strong first order one in $3d$
transition metal compounds.  It may be that the filling controlled
transitions are also qualitatively different in the strong SOC case.
There is very little theoretical work on this problem\cite{wang-senthil,You11,okamoto},
which promises to be a rich subject for future study, and not only in Sr$_2$IrO$_4$. 

Another important topic which has been mentioned but not given in
depth attention in this review is the emergence of fractionalized
exotic phases from topological bands.  This possibility is vetted
experimentally and theoretically by the most studied fractionalized
states of matter, the Fractional Quantum Hall Effect (FQHE) state of
two-dimensional electron gases in high magnetic fields.  The Landau
levels of that problem can be view as the simplest examples of bands
with non-trivial Chern number, and it is understood that the FQHE
originates from interacting electrons that partial fill a Landau
level.  Many theorists have suggested that FQHE may occur due
to fractionally filled bands with non-trivial Chern number in
crystalline materials with broken time-reversal symmetry: a fractional
Chern insulator\cite{sheng11,neupert11,tang}.  Time-reversal symmetric 
fractional topological insulators\cite{2D_frac_TI,3D_frac_TI,3D_frac_TI_senthil} have also been
proposed in both 2D and 3D, as have other fractionalized phases such as the topological
Mott insulator\cite{pesin,will-tmi}.
While current theoretical understanding demonstrates  
clearly that all these phases {\em may} exist, \emph{i.e.}\ that they are
stable phases of matter for some Hamiltonian parameters, they are all
outside the domain of conventional mean field approaches, including
those combined with {\em ab initio} methods such as LDA+U and DMFT.
For many of the other phases described in this review, however, mean
field approaches are adequate.

Also beyond these mean field approaches are numerous possible quantum
critical points (as well as thermal ones).  The rich variety
of phases found already in the studies discussed in this review
implies that suitable phase transitions between these phases should be
accessible.  This includes of course MITs, but
also transitions to the onset of magnetic order, and between different
types of band topology.  In many experimental systems with weak SOC,
quantum criticality is seen to coincide with unconventional
superconductivity, and many believe the former is a causative factor
in the latter\cite{QC}.  It will be very interesting to seek superconductivity
from quantum critical fluctuations in strong SOC materials, and if it
exists, to see whether topological superconductivity might result.
Experimentally, strong SOC tightens the link between electronic and
lattice structure, so that a very strong influence of pressure/strain
may be expected in these materials.  We anticipate a growing number of
studies of criticality in this class of systems.

As in any developing field, there are controversies.  The $J_{\rm
  eff}=1/2$ limit is actively debated in different iridates.  The
degree of spin-orbital entanglement can be addressed by several
distinct experiments, and should vary from material to material,
depending upon the strength of crystal fields, etc.  While the $J_{\rm
  eff}=1/2$ picture is appealing and simple, it is probably, however,
not essential to many of the phenomena we have discussed.  Also at
issue is the nature of the correlated insulating state.  Some authors
distinguish ``Mott'' and ``Slater'' insulators, where in the latter
case insulating behavior is due to magnetic order.  It is not clear
whether this distinction is well-defined, especially at zero
temperature, given that magnetic order itself can arise only from
Coulomb interactions. In general, there is no reason to have a phase
transition as the interaction strength is increased up to the local
moment regime, if the magnetic ordering pattern is unchanged.
Different experiments\cite{damascelli,hsieh2012observation} that try to differentiate the
two may be formulated: is the system conducting or insulating above
the magnetically ordering temperature?  Is the charge gap
$\Delta_c$ large compared to the exchange $J$?  Is the MIT 
continuous or first order?  These criteria need not agree.
A related issue is the ``degree of correlation'' in, \emph{e.g.}\ iridates,
with presumably Slater insulators being less correlated.  This
question is typically posed by {\em ab initio} based approaches
such as LDA+U or DMFT.  We remark that by these techniques, even the
correlation strength of high-T$_c$ cuprates might be
questioned\cite{PhysRevB.80.054501}, so the physical significance of
this question is opaque to us. Finally, the strong SOC limit itself is
open to question.  The strong SOC limit indeed depends not only upon
the ratio of SOC to non-cubic crystal fields, but also on the strength
of hopping or exchange relative to SOC.  Mazin {\em et al.}\ have
suggested indeed that SOC is not critical to the insulating state in
Na$_2$IrO$_3$\cite{foyevtsova2013ab,mazin2013origin}.

Moving beyond these debates, perhaps one of the most ambitious programs for
the future is ``engineering'' desired electronic states.
An appealing possibility is combining the materials discussed in this
review with advances in layer by layer atomic growth of oxides to
produce designer heterostructures of iridates and other strong SOC
systems. For example, various heterostructures of perovskites\cite{xiao2011,wang_ran13},
(including SrIrO$_3$) as well as of $R_2$Ir$_2$O$_7$ along the [111]  
direction\cite{hu_fiete13} have been proposed and several non-trivial
phases such as topological insulators and quantum Hall states are
predicted to exist under certain conditions.  Experiments of this type 
are within the realm of possibility in the near future.  

At the time of writing, the materials synthesis and experimental
characterization of diverse 4$d$ and 5$d$ transition metal oxides is
only increasing. The effects
of carrier doping, hydrostatic pressure, and high magnetic fields are
under investigation.   The outcome of these experiments and existing
and future theory may uncover entirely new classes of many-body
quantum ground states in strongly spin-orbit coupled systems.

\section*{Acknowledgments}
We thank S.~Bhattacharjee, A.~Burkov, R.~Chen, A.~Go, M.~Hermele, G.~S.~Jeon, E.~K.~H.~Lee, E.-G.~Moon, D.~Pesin, 
K.~Park, L.~Savary, C.~Xu, and B.-J.~Yang for collaboration on work related to this review, and for helpful
discussions. We are also grateful to G.~Cao, H.~D.~Drew, P.~Gegenwart, B.~J.~Kim, Y.-J.~Kim, S.~R.~Julian,~S. Nakatsuji, R. Perry,
Y. Singh, A.~B.~Sushkov, and H.~Takagi for sharing their experimental data and fruitful discussions.
Our special thanks go to J. J. Ishikawa, S. Nakatsuji, and D. E. MacLaughlin for providing the
plot used in Figure \ref{iridate_data} (a).
This work was supported by Walter Sumner Foundation (WWK), DOE award No.~DE-SC0003910 (GC), NSERC,
CIFAR, Center for Quantum Materials at the University of Toronto (YBK),
and NSF grant No. DMR--1206809 and DOE Basic Energy Sciences Grant
No. DE-FG02-08ER46524 (LB).  Some of this work was carried out at the
KITP, funded by NSF grant PHY--1125915. Research at Perimeter Institute 
is supported by the Government of Canada through Industry Canada and by the 
Province of Ontario through the Ministry of Research \& Innovation.


%

\end{document}